\documentclass[aps,pre,
twocolumn,
superscriptaddress,
floatfix,
amsfonts,
amsmath,
amssymb,
citeautoscript]{revtex4-1}
\usepackage{times}
\usepackage{graphicx}
\usepackage{dcolumn}
\usepackage{bm}
\usepackage{color}
\usepackage{amssymb}  
\usepackage{amsmath}   
\newcommand {\beq} {\begin{equation}}
\newcommand {\eeq} {\end{equation}}
\newcommand {\bqa} {\begin{eqnarray}}
\newcommand {\eqa} {\end{eqnarray}}

\usepackage{hyperref}

\begin{document}
\sloppy

\title{Significance of the nature of disorder on the universal features of the spatio-temporal correlations of two-dimensional Coulomb-clusters}
\author{Jami Prashanti}

\affiliation{Department of Physical Sciences, 
Indian Institute of Science Education and Research Kolkata,
Mohanpur - 741246, WB, India }

\author{Biswarup Ash}
\affiliation{ Department of Physics of Complex Systems, 
Weizmann Institute of Science, Rehovot 7610001, Israel}

\author{Amit Ghosal}

\affiliation{Department of Physical Sciences, 
Indian Institute of Science Education and Research Kolkata,
Mohanpur - 741246, WB, India }

\date{\today}

\begin{abstract}

We comprehend the role of imperfections in materials consisting of interacting particles, arising from different origins on their universal features. Specifically, we report the static and dynamic responses in a cluster of Coulomb interacting particles in two dimensions. Confined systems with pinned impurities are studied, and results are compared with those from irregularly trapped system of particles. While the disorder of first type leads to diffusive single-particle dynamics, the motion of a single particle in an irregular trap is chaotic but ballistic. The many-particle system does not differentiate between these two models of disorder insofar as their qualitative properties are concerned -- particularly for describing the thermal melting of the underlying Coulomb-``solid". However, quantitative differences persist -- the relaxation time scales differ significantly by tuning impurity concentration.
\end{abstract}

\maketitle
\section{Introduction}
The effect of quenched disorder on the properties of systems, which would be otherwise ordered, has been a subject of extensive study~\cite{pp1,PhysRevLett.121.205501,ig,C3SM50998B,PhysRevLett.126.178001,bir1,pnas,Szamel_2013,PhysRevLett.104.015701}. In particular melting transitions in two-dimensions (2D) draws special attention because the translational order and bond-orientational order set in at different threshold temperatures. This yields a hexatic phase between the solid and liquid states in clean 2D systems~\cite{RevModPhys.60.161,KT1,KT2,HN1,HN2,Young}. The role of disorder on such a melting, as well as the fate of the hexatic phase under the influence of disorder has generated significant research attention~\cite{PhysRevLett.114.035702,PhysRevLett.107.155704}. Such studies are relevant to physical systems under experimental investigations~\cite{PhysRevE.77.041406,PhysRevLett.127.018004,PhysRevResearch.4.023116}. These include randomly pinned Abrikosov vortex lattice in type-II superconductors~\cite{ABRIKOSOV1957199}, Wigner crystals~\cite{PhysRev.46.1002,PhysRevB.49.2667,rev1}, dusty plasma~\cite{Bonitz_2010,PhysRevB.20.326,Bonitz_2008,Melzer_plasma_Exp,PhysRevResearch.4.023116,PhysRevLett.72.4009}, among others. Different systems are disordered by different means, and often times the microscopic origin of disorder remains unknown. Still, it is a conventional wisdom that there are universal qualitative features of disordered systems which are not sensitive to the microscopic details of the underlying impurities.

Unfolding of the complex interplay of inter-particle interactions and disorder has generted important research ideas in the broad area of condensed matter physics. Unraveling such an  interplay in small systems with finite number of particles is significant from both technological and fundamental physics perspectives~\cite{PhysRevLett.74.458,Kouwenhoven_1998,Bonitz_2008,PhysRevLett.82.3364,Simulation,PhysRevE.95.042603}. While a sharp phase transition cannot be expected in a confined system~\cite{PhysRev.87.404}, a thermal crossover from a \lq solid \rq-like to a \lq liquid \rq-like phase still occurs~\cite{Ghosal2006}.

How significant are the microscopic details of disorder for properties of system of interacting particles? For example, disorder can arise from the randomness of the pinning centers in a system. In such a model of disorder, a single test-particle travels diffusively. On the other hand, an irregularity in the trapping potential can also make the trajectory of the test-particle unpredictable, leading to a ballistic yet chaotic motion. Is this fundamental difference in the motional signature of a single particle of any relevance for describing the universal properties of a disordered system composed of many interacting particles? We address this question in this article considering (long-range) Coulomb interacting particles in confined geometries.

Transport in small grains had attracted significant research interests in the past ~\cite{PhysRevLett.91.246801,RevModPhys.79.469}. Such a confined system, when the confinement breaks all spatial symmetries, behaves like a disordered system. Such systems are expected to show universal statistics of energy levels and energy functions~\cite{MIRLIN2000259}. An interesting comparison of chaotic versus diffusive dynamics has been carried out in experiment using microwave studies~\cite{doi:10.1080/09500340210145286}. An important difference between disordered systems with diffusive and chaotic dynamics, however is the following: A strong pinning disorder can lead to localization defying the applicability of random matrix theory~\cite{RevModPhys.72.895}, but a ballistic motion is extended across the whole system\cite{PhysRevLett.77.4744}. Nevertheless, for weaker disorder strengths a comparison between the motions of interacting many-particles in different environments is an intriguing point to explore.
  
Addressing these issues, our key results are the following: The systems with pinned impurities possess a more robust positional order than in irregular confinements, while the bond-orientational order responds in a similar manner in both systems. 
The details of the disorder have little role on the ``melting" of the Coulomb ``crystal"~\cite{Ghosal2006}. Both the systems show qualitatively similar motional signatures. However, they differ in quantitative details. The pinning disorder increases the relaxation time considerably, in agreement with existing literature~\cite{Kim_2003}.
The rest of this article is organized as follows. Section II explains our model and its parameters and the details of our numerical simulations. Section III discusses different static properties in different subsections. We also estimate the thermal crossover from a solid-like to a liquid-type behavior from the temperature ($T$) dependence of the positional and bond orientational orders. In section IV, we calculate the dynamic observables focusing on the temporal evolution of these two differently disordered systems. Finally, Section V describes the one- and two-particle dynamics in terms of our analysis of Poincare sections in the disordered systems. We conclude our results in section VI.

\section{Model and Method}
We consider a system of $N$ classical particles in a confining potential. These particles interact with Coulomb repulsion and are restricted in a two-dimensional (2D) $xy$-plane. We neglect screening in our small finite clusters~\cite{bhattacharya2013melting}. The Hamiltonian for a circular confinement can be given by,
\begin{equation}\label{e1}
H=\sum_{i<j=1}^{N}\frac{1}{\left|\vec{r}_{i}-\vec{r}_{j}\right|}+\sum_{i}r_{i}^2 ,
\end{equation}
Here, $r_{i}= \left|\vec{r}_{i}\right|=\sqrt{x_{i}^2+y_{i}^2}$ is the distance of $i$-th particle from the center of the confinement. The first term describes the Coulomb repulsion of particles. The second term represents the confinement potential.
In writing the Hamiltonian in Eq.~\eqref{e1} we used a rescaled variables $r'\rightarrow\phi^\frac{1}{3}\alpha^\frac{-1}{3} r$,    
   and energy $E'\rightarrow\phi^\frac{2}{3}\alpha^\frac{1}{3}E$, where $\phi=\frac{q^2}{\epsilon}$ and $\alpha=m\omega_{0}^2$
  where $\omega_{0}$ sets the average density of particles in the system, so that the original Hamiltonian in the primed coordinate has the standard form\cite{PhysRevB.49.2667,PhysRevE.96.042105}: 
\begin{equation}\label{e2}
H=\frac{q^2}{\epsilon}\sum_{i<j=1}^{N}\frac{1}{\left|\vec{r}_{i}-\vec{r}_{j}\right|}+ \sum_{i}^{N}V_{\rm conf}^{\rm cr}(r_{i}) .
\end{equation}
where,
\begin{equation}
 V_{\rm conf}^{\rm Cr}(r)=\frac{1}{2}m\omega_{0}^2r^2 ,
\end{equation}
 In order to represent a ``clean" and regular system in 2D, we choose the $ V_{\rm conf}^{\rm Cr}(r)$, in which the classical motion of a single particle is integrable.\cite{PhysRevB.49.2667} The ``cleanliness" of the regular circular potential is best understood when we compare our results with those for disordered traps.
 
As the first example of disordered system, we introduced ``pinning" in the aforementioned $V_{\rm conf}^{\rm Cr}(r)$. Pinning is introduced by quenching a fraction $(n_{\rm imp})$ of the constituent particles in random location within the trap. Disorder averaging~\cite{PhysRevLett.90.176801} is performed by taking 12 independent realizations of impurity configurations for each $n_{\rm imp}$. Such a pinning disordered system is expected to show a universal features of diffusive dynamics of particles in the trap. Note that here $n_{\rm imp}$ controls the strength of disorder.

We considered the following model of irregular trap as the second model of our disordered system:
\begin{equation}\label{e3}
V_{\rm conf}^{\rm Ir}(r) = a\left[ \frac{x^4}{b}+by^4-2{\lambda}x^2y^2+{\gamma}(x-y)xyr \right].
\end{equation}
Here, $V_{\rm conf}^{\rm Ir}(r)$ breaks all spatial symmetries and is also known to generate ballistic and chaotic motion of a single particle within it \cite{BOHIGAS199343,bhattacharya2013melting}.
 Together with the lack of spatial symmetries, the chaotic motion of the single particle is taken to be a signature of disorder in the irregular trap.   
This system is characterized by four parameters $a$, $b$, $\lambda$, and $\gamma$ \cite{PhysRevE.98.042134}, which could be tuned to adjust the disorder strength. 
The overall multiplicative factor $a(> 0)$  controls the average density of particles in the system without effecting the trajectory of a single particle in the confinement.
The values of other parameters used are  $b=\pi/4, \lambda = [0.565,0.635]$ and $\gamma = [0.10,0.20]$ \cite{PhysRevLett.90.026806}. We generate upto 7 independent realizations of the irregular confinement.

By rescaling variables as before and $a=(m\omega_{0}^2/2r'^2)a'$, the Hamiltonian for an irregular confinement reads as,
\begin{eqnarray}\label{e4}
H &=& \sum_{i<j=1}^{N}\frac{1}{\left|\vec{r}_{i}-\vec{r}_{j}\right|}+
\sum_{i}a'\left[ \frac{x_{i}^4}{b}+by_{i}^4-2{\lambda}x_{i}^2y_{i}^2 \right. \nonumber \\
  &+& \left. {\gamma}(x_{i}-y_{i})x_{i}y_{i}r_{i}\frac{}{} \right].
\end{eqnarray}
The time scale is renormalized to $t'=\hbar\phi^\frac{-2}{3}\alpha^\frac{-1}{3}t$. Such re-normalization of length, energy, and time eventually expresses the temperature as $T'=\frac{E'}{k_{B}}$, where $k_{B}$ is the Boltzmann constant.

In order to explore the universal features of disordered systems we aim to compare the role of the diffusive dynamics in the pinned system versus the chaotic and ballistic motion of Coulomb particles in the irregular trap.
In a confined geometry at $T=0$, the ground state will be the one that is consistent with the symmetry (or the lack of it) of the underlying trap.
Upon increasing $T$, the constituent particles of the Coulomb cluster execute random excursions from their equilibrium position, and beyond a threshold temperature, $T_{c}$ the crystal ``melts"\footnote{A bulk system shows diverging susceptibility at a transition which turns into a broad peak at $T_{X}$ at the thermal crossover in finite systems}. Such melting in the bulk systems are tuned by the dimensionless parameter $\Gamma=\sqrt{n\pi}/T$ ~\cite{PhysRevB.20.326}. For our comparative study of clean and disordered traps, we induce melting by tuning $T$ alone, and we consider systems which are composed of the same total number of particles $N$, and have the same density $n$ of particles. This is adjusted by appropriately tuning the parameter $a^{\prime}$ in Eq.~\eqref{e4}.

To study the physical observables, we carried out the classical molecular dynamics (MD)~\cite{frenkel2001understanding} with $N=150$ particles.

We performed MD runs for up to $10^7$ MD steps with a time step size of $\delta t = 0.005$ in our dimensionless unit. 
For a fair comparison, the strength of disorder was matched by comparing the bond-orientational order (BOO) as discussed in the next section.

Solidity in a clean 2D system in the thermodynamic limit is associated with long-range orientational order and quasi long-range positional order \cite{PhysRevLett.17.1133}. In what follows, we explore the pair correlation function and 6-fold BOO in order to gain insight into the thermal melting of Coulomb clusters in traps.

\section{Static Properties}
\subsection{Bond Orientational Order}
 The Bond orientational order (BOO) of a system having 6-fold symmetry like for a triangular lattice is defined as~\cite{nelson2002defects},
\begin{equation}\label{e5}
\psi_{6}(r_k)=\frac{1}{N_{b}}\sum_{l=1}^{N_{b}}
\exp{(i6\theta_{kl})}.
\end{equation}
Here, the angle $\theta_{kl}$ is the bond angle between the particle $k$ and with its nearest neighbors $l$ about an arbitrary (but fixed for all nearest neighbors) axis.
We identify the nearest neighbors of particles by using the Voronoi construction~\cite{TIPPER1991597}. For the purpose of a fair comparison, we wish to find $n_{\rm imp}$ that typifies the strength of disorder in aforementioned irregular system. This is implemented by studying the distribution of $|\psi_{6}|$ from the pinned circular trap with different $n_{\rm imp}$'s and then identifying the one that produces a $P(|{\psi_{6}}|)$ closest to the one generated from the $V_{\rm conf}^{\rm Ir}(r)$ for the same number of particles participating in the dynamics at the same $T$.
\begin{figure}[t!]
\includegraphics[width=0.46\textwidth]{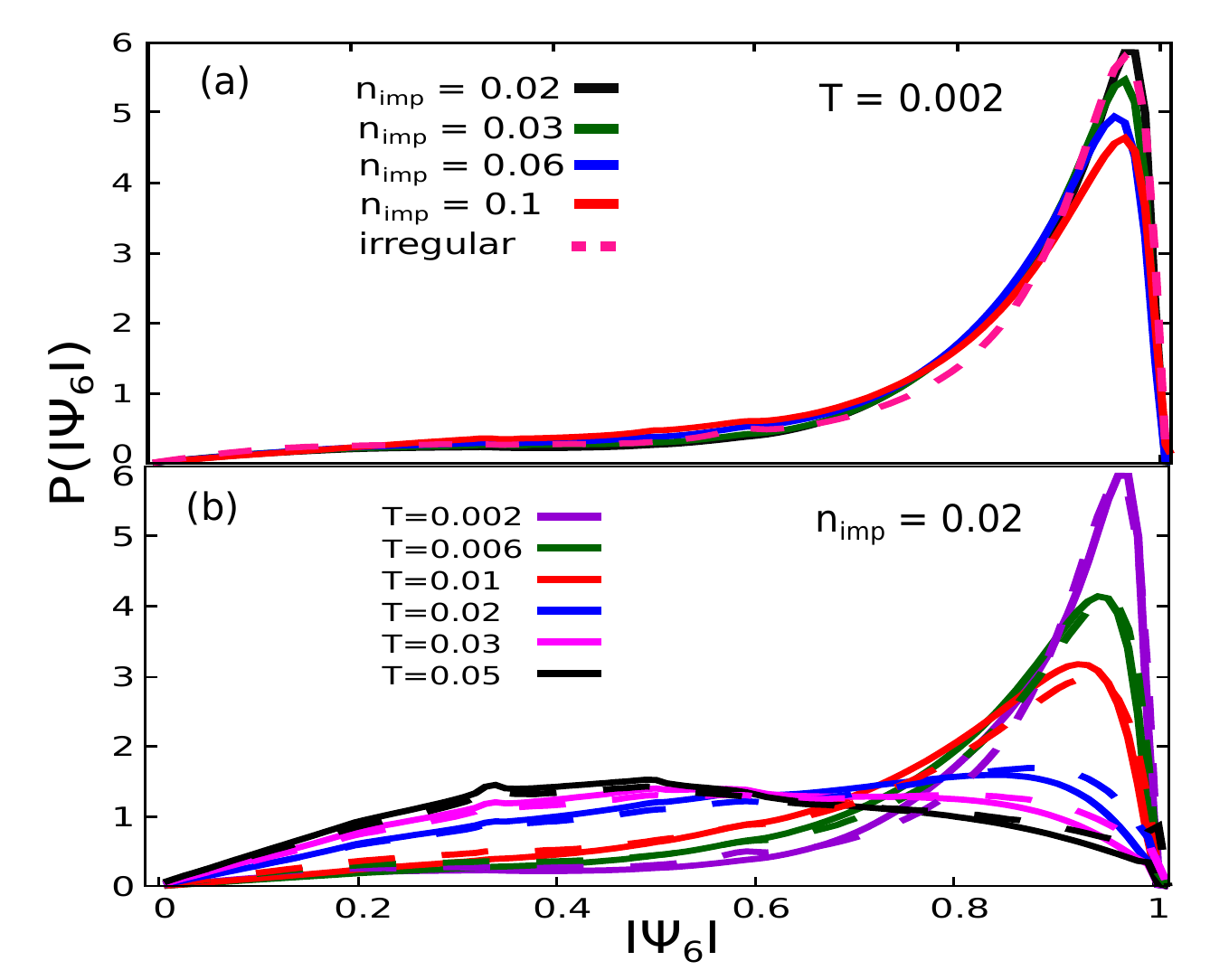}
\caption{(a) The distribution $P(|{\psi_{6}}|)$ at $T=0.002$ is shown by color traces for different $n_{\rm imp}$. The dotted line represents the $P(|{\psi_{6}}|)$ for irregular confinement at the same $T$. Evidently, the dotted line matches best with the results in pinned system for $n_{\rm imp}=0.02$. (b) The thermal evaluation of $P(|{\psi_{6}}|)$ in both the pinned circular system ($n_{\rm imp}=0.02$, solid lines) and irregular (dotted lines) indicates that the $T$-dependence of $P(|{\psi_{6}}|)$ is nearly identical.  }

\label{f1}
\end{figure}

Fig.~\ref{f1}(a) demonstrates the dependence of $P(|{\psi_{6}}|)$ on the pinning concentrations $n_{\rm imp}$ at $T=0.002$ -- the lowest temperature of our study. We find that for low $n_{\rm imp} \in [0.02$-$0.1]$ shown in Fig.~\ref{f1}(a), $P(|{\psi_{6}}|)$ alters only weakly and the trace for $n_{\rm imp}=0.02$ finds the best match of $P(|{\psi_{6}}|)$ with that obtained from $V^{\rm Ir}_{\rm conf}(r)$ (shown by the dashed line). 

Subsequently, we study the effect of $T$ on $P(|{\psi_{6}}|)$ for $n_{\rm imp}=0.02$ in Fig.~\ref{f1}(b). Our results demonstrate that the temperature evolution of $P(|{\psi_{6}}|)$ in the range $T=0.002$-$0.05$ for the pinned and irregular confinements is nearly identical.

In the limit $T \rightarrow 0$, a perfect triangular lattice in the bulk system yields $P(|{\psi_{6}}|)$ a $\delta$-function at ${\psi_{6}} = 1$. A confined system develops a tail in $P(|{\psi_{6}}|)$ for $|{\psi_{6}}| \leq 1$ even at $T=0$,\cite{PhysRevE.91.032312,PhysRevE.67.021608} but the strong peak at ${\psi_{6}} = 1$ persists. $P(|{\psi_{6}}|)$ changes from a sharply peaked distribution at low $T$ to a broad, and nearly symmetric distribution for a high $T$ melted state in confinements~\cite{bhattacharya2013melting}. Thus we conclude that the diffusive versus chaotic-ballistic dynamics of single particles in the respective traps have little effect on the thermal evolution of the bond-orientation ordering of Coulomb-interacting particles in those confinements.

\subsubsection{Orientational susceptibility}

While the $T$-dependence of $P(|{\psi_{6}}|)$ depicts the thermal crossover from a solid-like phase to a liquid-like phase at a quantitative level, it is the study of orientational susceptibility which estimates the crossover temperature scale $T_{X}$ which we discuss in the following. The orientational susceptibility~\cite{Sun2016} is defined as,
\begin{equation}\label{e6}
\chi_{6}=N[\langle \left|\psi_{6}\right|^2\rangle-\langle \left|\psi_{6}\right| \rangle^2].
\end{equation}  
Here, the angular brackets $\langle \cdots \rangle$ implies average over all particles, as well as over MD configurations at a given $T$. 
\begin{figure}[t!]
\includegraphics[width=0.46\textwidth]{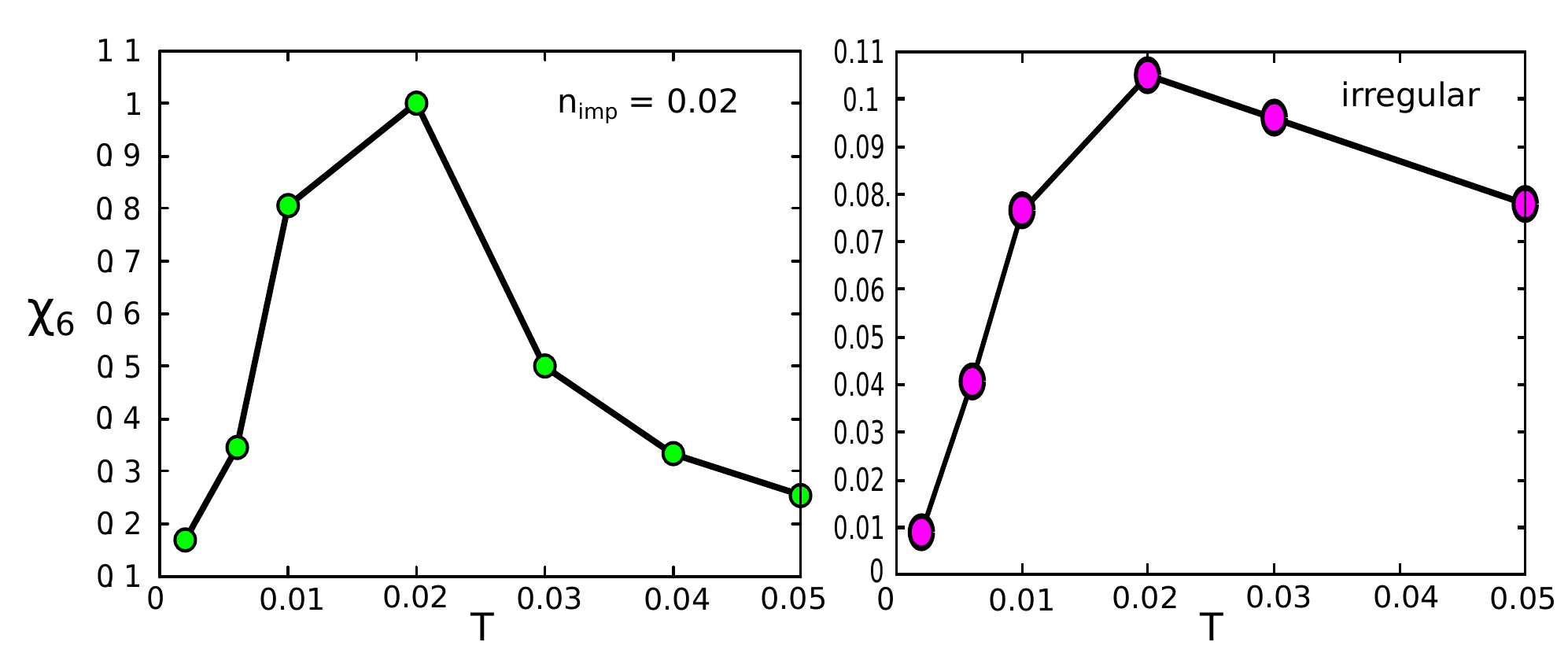}
\caption{Temperature dependence of the fluctuations in bond orientational order shows a hump in $\chi_{6}(T)$ around $T_X\approx 0.02$ in both pinned circular ($n_{\rm imp}=0.02$) and irregular confinements, though the peak in $\chi_{6}(T)$ is sharper for pinned system.}
\label{f3}
\end{figure}
The temperature dependence of $\chi_{6}$ for pinned and irregular traps is shown in Fig.~\ref{f3}. The broad hump in $\chi_{6}(T)$ centered around $T_X \approx 0.02$ is discernible in both the traces.
Note that the actual values of $\chi_{6}$ are different in the two different models of disorders, though the location of the hump at  $T_{X}$ is very similar. This emphasizes our earlier conclusion that the diffusive versus chaotic-ballistic dynamics of particles in the respective traps has a minor role on the melting of Coulomb-interacting particles in these confinements.

Having studied the thermal evolution of the orientational ordering on different traps, we next proceed to explore the positional order in these systems.

\subsection{Pair Correlation Function}
Positional order in a system is tracked by the pair correlation function $g(r)$, which yields the probability of finding a particle at a distance $r$ on an average, from a reference particle, and is defined as~\cite{mcdonald2006theory}
\begin{equation}\label{e7}
g(r)=\frac{1}{2\pi rN}
\sum_{i=1}^{N}\sum_{j\neq i=1}^{N}
\langle \delta(r-|\vec{r_{i}}-\vec{r_{j}}|) \rangle .
\end{equation}

\begin{figure}[t!]
\includegraphics[width=0.46\textwidth]{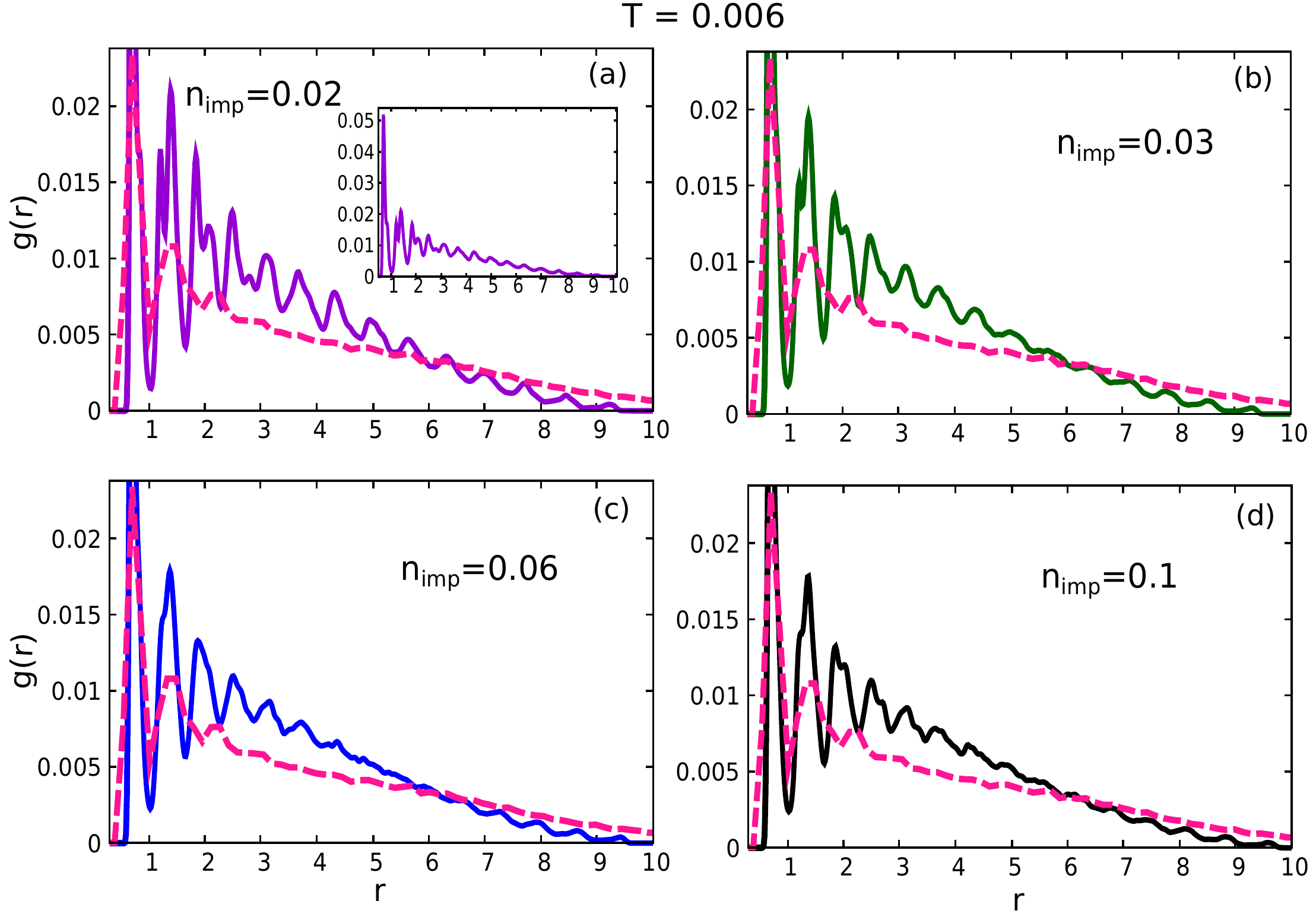}
\caption{The solid lines show $g(r)$ at $T=0.002$ in pinned circular systems for different $n_{\rm imp}$ (in different pannels), while the dotted line is the result from $V_{\rm conf}^{\rm Ir}(r)$. These imply that a healthy positional order persists in pinned system in comparison with that in $V_{\rm conf}^{\rm Ir}(r)$. The height of the first peak of $g(r)$ has been chopped-off from all panels for visual clarity, though a representative plot of entire $g(r)$ is shown as an inset of panel (a).}
\label{f4}
\end{figure}

\begin{figure}[t!]
\includegraphics[width=0.46\textwidth]{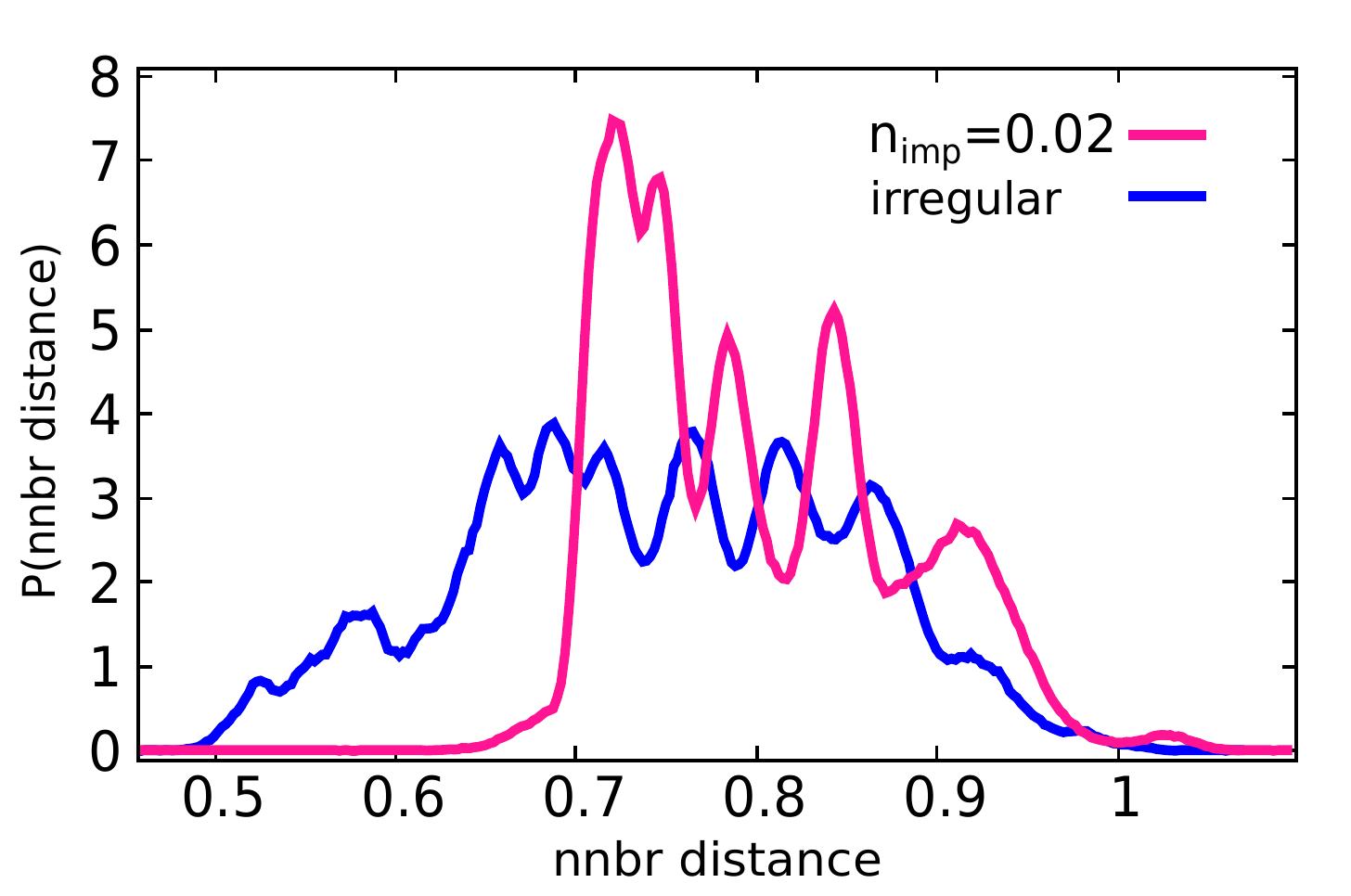}
\caption{The distribution of the nearest neighboring distances of pinned circular (pink) and irregular (blue) traps illustrate that the such distances are more inhomogeneous in irregular trap and thereby weaken the positional order more easily compared to that in pinned system, as found in Fig.~\ref{f4}(a).}
\label{f5}
\end{figure}

\begin{figure}[t!]
\includegraphics[width=0.46\textwidth]{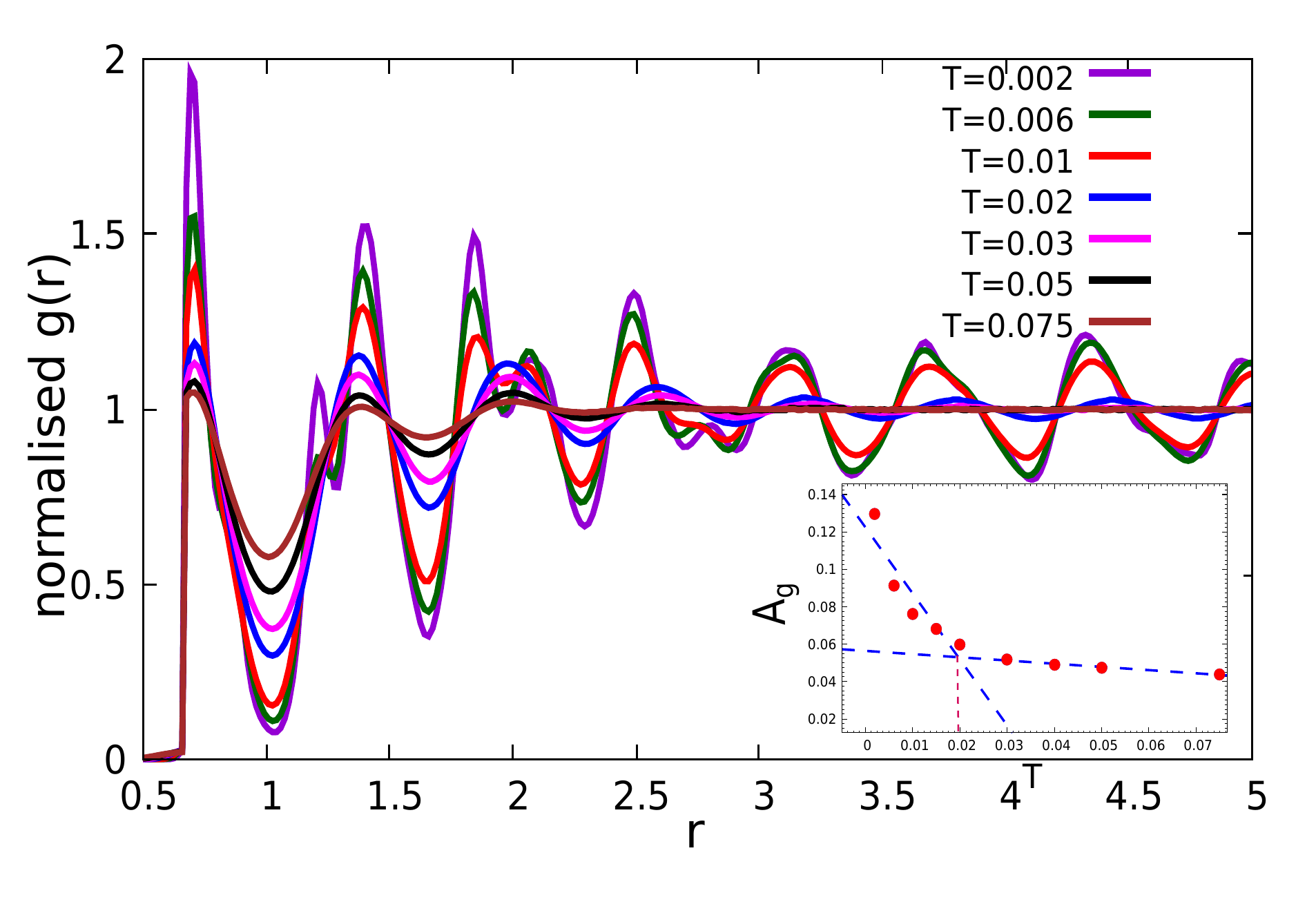}
\caption{Thermal evolution of the spatially modulated part of $g(r)$ (after removing the smooth r-dependence) for $n_{\rm imp} = 0.02$. The modulations weaken with $T$ and signal a crossover to a liquid-like state. The magnitude of the total area $A_{g}$ under each trace (disregarding their sign) decreases with $T$. The decrease of $A_{g}$ with $T$ is shown in the inset, which show two nearly linear branches for small- and large-$T$. The change of slope occurs at $T_X\approx 0.02$.}
\label{f7}
\end{figure}

In Fig.~\ref{f4}(a-d) we demonstrate the dependence of $g(r)$ at low $T$ for $n_{\rm imp} = 0.02, 0.033, 0.06$ and $0.10$ respectively using solid lines. The dashed line represents $g(r)$ calculated for $V^{\rm Ir}_{\rm conf}(r)$. The long-range nature of the decaying envelope of oscillations in $g(r)$ measures the positional order in the systems. With the increase of $n_{\rm imp}$, such oscillation becomes progressively damped, yet,
it is evident that the pinned system displays a robust positional order compared to the irregular system. It is also interesting to note that unlike for the case of $P(|{\psi_{6}}|)$, the nature of $g(r)$ in the system with pinning disorder does not find a reasonable match with that for $V_{\rm conf}^{\rm Ir}(r)$ in the entire range of $n_{\rm imp} \in [0.02$-$0.10]$. Thus the diffusive versus chaotic-ballistic motion of single particles in the respective traps indeed differ in displaying positional order. In the following we proceed to comprehend the origin of this difference. \\
\hspace*{5mm}
In pinned system the symmetry of a six-coordinated neighborhood (just the existance, and not the quantitative measure) gets disturbed only in a small local region around the pinning centers, leaving out the configuration of other particles largely unaltered elsewhere in a circular trap. On the other hand, the spatial undulations of $V^{\rm Ir}_{\rm conf}(r)$ persist over the entire region of the trap, which causes distortions of the lattice lines everywhere in the system of the underlying triangular lattice. As a result, positional order weakens without disturbing the orientational order in $V^{\rm Ir}_{\rm conf}(r)$. To illustrate this further, we present the distribution of the nearest neighbor's distances in the two confinements in Fig.~\ref{f5} which supports the aforementioned argument in the following manner.
 Fig.~\ref{f5} presents the distribution of the nearest neighbor distances in the pinned system ($n_{\rm imp}=0.02$) and for irregular trap at $T=0.002$. The nearest neighbors of a particle is identified using Voronoi construction~\cite{TIPPER1991597}. The distribution makes it evident that the nearest neighbor distances occur over a wider range of values in $V_{\rm conf}^{\rm Ir}(r)$, which washes out the periodic oscillations in $g(r)$ and thus weakens the positional order,

 compared to the pinned circular trap. \\
\begin{figure}[t!]
\includegraphics[width=0.46\textwidth]{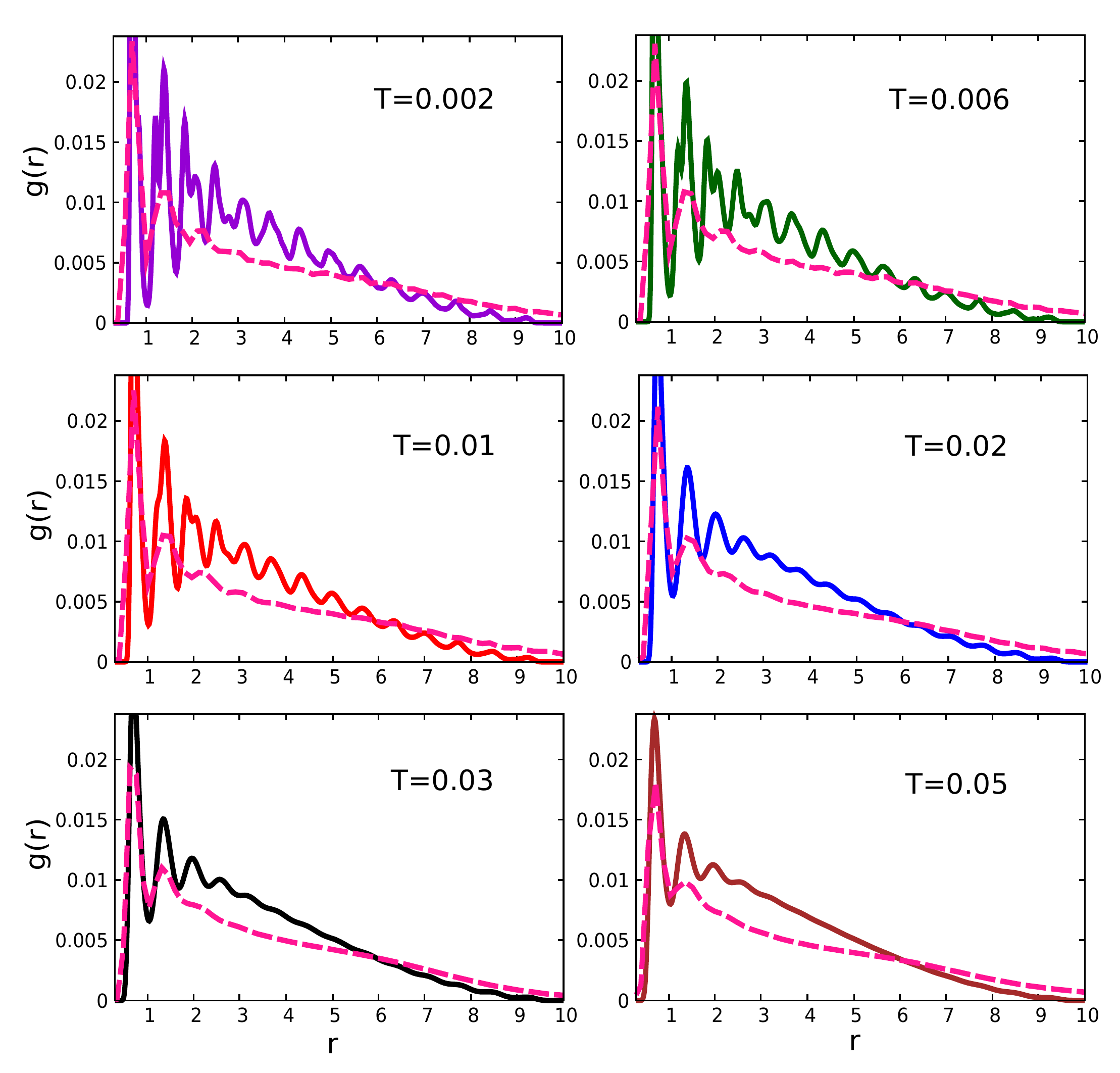}
\caption{Thermal evolution of $g(r)$ shown for $n_{\rm imp} = 0.02$. The results from irregular trap is also shown by the dotted line for comparison. It is evident from the figure that the pinned system supports a more robust positional order (as also found in Fig. 3)}
\label{f6}
\end{figure}
Having established the weaker dependence of positional order on $n_{\rm imp}$, at least in the range of $n_{\rm imp}$ studied, we proceed to explore the thermal evolution of $g(r)$ which is shown in Fig.~\ref{f6}. With increasing $T$, the initial peaks and the oscillations of $g(r)$ start fading away from its tail.
The irregular system (dashed line) shows only a strong first peak followed by a few feeble humps in the $g(r)$, indicating depleted positional order even at the lowest $T$ -- a characteristic of a liquid-like state~\cite{PhysRevE.96.042105}. Upon increasing $T$, the height of the initial peaks decreases, and all higher-order modulations
disappear beyond  $T > 0.020$.
 Note that in our study $g(r)$ goes to zero for large $r$ due to the finiteness of our system. This is unlike the bulk system for which $g(r \rightarrow \infty)=1$ \cite{pathria2016statistical}. For estimating a $T_{X}$ from $g(r)$, we employ the following procedure. In order to amplify the effect of the modulation of $g(r)$ in traps, we eliminate the smoothly falling part of the $g(r)$.~\footnote{The smooth curve is obtained by repeated averaging of the falling part of $g(r)$ and subsequently, upon dividing the oscillatory falling part of $g(r)$ with the obtained averaged curve.} This resolves its oscillatory part and also allows it to saturate to the unit value at large $r$ like in bulk system, as shown in Fig.~\ref{f7}. Finally, the solid to liquid crossover is quantified by the total area $A_{g}$, under the modulating part of $g(r)$. As expected, the solid will have a larger $A_{g}$ form stronger and long-range oscillations, and $A_{g}$ must decrease with increasing $T$. The inset of Fig.~\ref{f7} shows that the nature of the fall of $A_{g}(T)$ has two near-linear branches, and we identify $T_X$ where the change of the slope takes place. Analyzing our data we obtain $T_{X} \approx 0.0195$. This is consistent with the previously estimated $T_{X}=0.019\pm 0.002$~\cite{PhysRevE.98.042134} from orientational susceptibility in an irregular trap. This is interesting to note that even though the pinned system has a more robust positional order, the estimate of $T_X$ remains insensitive. \\

Having performed the comparative analysis between the thermal crossover between a solid-like to a liquid-like state in different confinements, we conclude that the nature of the disorder and hence the underlying single particle dynamics has no significant role on the universal physics of disorder on `melting', though differences exist in the details. Next, we turn to analyze dynamical characteristics associated with melting of Coulomb clusters in these confinements.
\section{Dynamic Properties}
In an earlier study~\cite{PhysRevE.98.042134} it was shown that the Coulomb particles in irregular confinements generate intriguing motional signatures akin to glassy dynamics. Here we carry out studies of the time evolution of particles in the pinned circular trap to investigate if and how the nature of disorder affects the temporal properties. 
\subsection{Mean Square Displacement}
We start discussing our results by considering dynamical mean squared displacement (dMSD), which is a measure of the deviation of the position of a particle with respect to a reference position over time t:
\begin{equation}\label{e8}
\langle \Delta r^2(t) \rangle = \frac{1}{N}\sum_{i=1}^{N} \langle (\vec{r}_{i}(t)-\vec{r}_{i}(0))^2 \rangle.
\end{equation}
\begin{figure}[t!]
\centering
\includegraphics[width=0.46\textwidth]{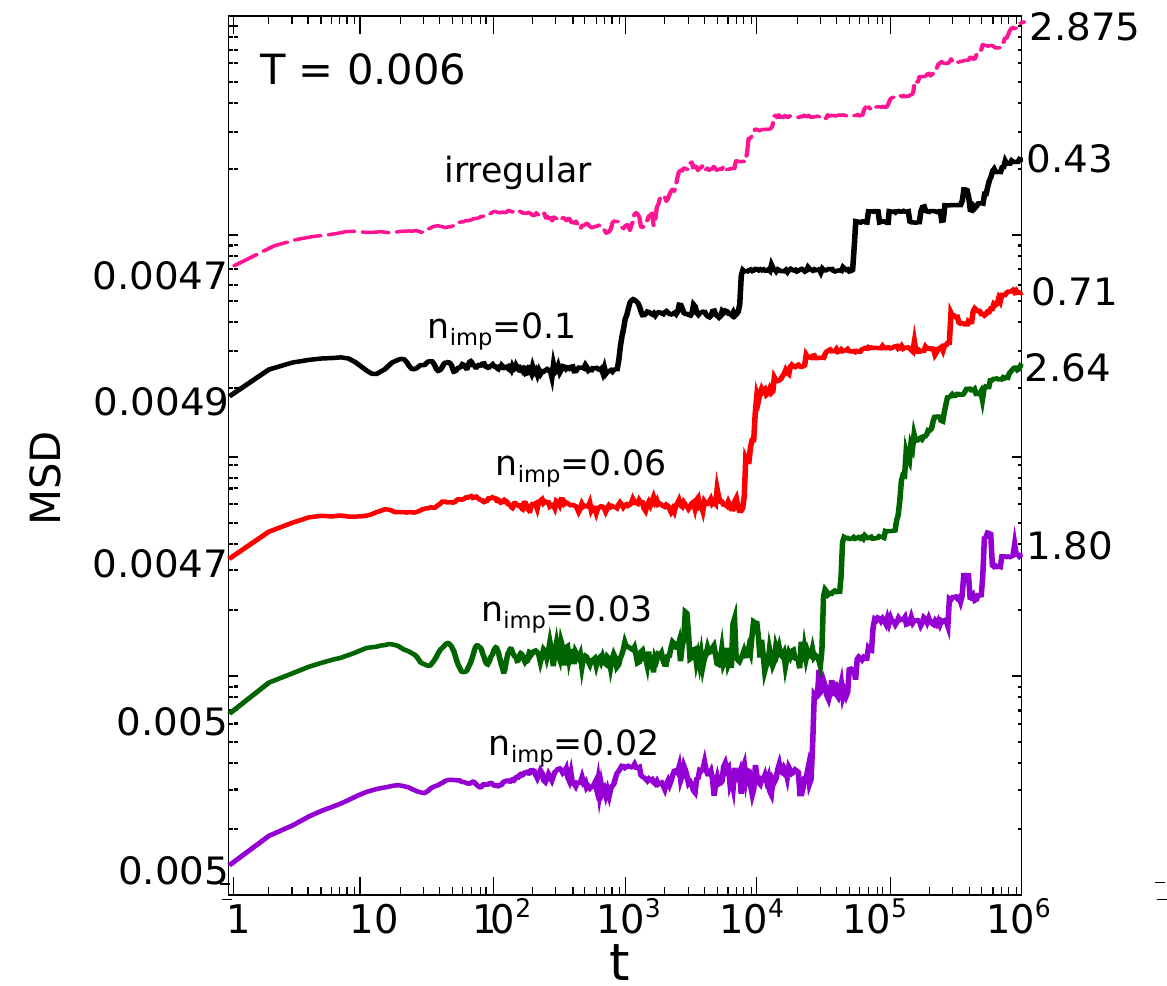}
\caption{Temporal evaluation of MSD at low $T$ ($=0.006$) for both irregular trap (dashed line) and pinned system (solid lines) with different $n_{\rm imp}$. The values of MSD for the largest $t=10^6$ are listed on the right $y$-axis for each trace. The results show an initial short time rise of MSD due to ballistic motion until the particles start feeling each other's presence. This is followed by a long plateau in MSD, and finally, its increase by occasional jumps to the next plateau at large times. Such a combination of `plateau-jump' structures in MSD is indicative of the motion of particles by cage-breaking events. We note that the "caging region'' (i.e., the plateaus in MSD) is longer for increasing $n_{\rm imp}$.}
\label{f8}
\end{figure}
We present in Fig.~\ref{f8} the trace of $\langle \Delta r^2(t) \rangle$ at low $T$ (in solid-like phase, $T=0.006$) for $n_{\rm imp}=0.02, 0.033, 0.06$ and $0.10$. The trace of $\langle \Delta r^2(t) \rangle$ obtained for $V^{\rm Ir}_{\rm conf}(r)$ is also shown as the dashed line.
The dMSD shows wide plateaus, separated by sudden jumps (even after averaging over MD configurations), as shown in Fig~\ref{f8}. 
Such a behavior arises from the ``caging effects"~\cite{BENNEMANN1999217,PhysRevLett.80.2338,PhysRevLett.98.188301} in which at low $T$, most of the particles remain localized for a long time in the cage formed by the nearest neighbors (on an average). This yields in wide plateau in the dMSD. The cage breaking occurs sporadically, letting a macroscopic fraction of particles, which are related to one another by nearest neighbor relationship, to move nearly at once, causing the jump in $\langle \Delta r^2(t) \rangle$ between successive plateaus. Such coherent cage breaking occurs because breaking of one cage typically moves the originally caged particle to the neighboring cage, and thereby making it unstable, which breaks this neighboring cage too~\cite{PhysRevE.98.042134}. And the process continues leading to a cascading effect.
Such a motional signature in steps causes dynamical heterogeneities~\cite{2011}, which in turn leads to a glassy dynamics~\cite{PhysRevE.98.042134,berthier2011dynamic}. As demonstrated in Fig.~\ref{f8}, we found aforementioned qualitative motion is insensitive to the nature of the disorder in the confinements, however, quantitative differences exist. For example, we found that the cage relaxation time-scale \cite{PhysRevE.96.042105} is typically much shorter ($\sim 10^2$) in irregular trap, whereas it is much larger ($\sim 10^4$) in our pinned circular system. This is expected, because the pinned particles prohibit cooperative rearrangements \cite{Kim_2003}, which increases the relaxation time. Also, the Fig.~\ref{f8} shows that the increase of $n_{\rm imp}$ shortens the time-scale for the cage relaxations, as expected. The more localized nature of the particles with pinning disorder (than in case of $V_{\rm conf}^{\rm Ir}(r)$) is also evident from the dMSD results, which shows that the pinning restricts the particle from traveling a significant distance as seen from Fig.~\ref{f8}.
\subsection{Van Hove Correlation function}
\begin{figure}[t!]
\centering
\includegraphics[width=3.5in]{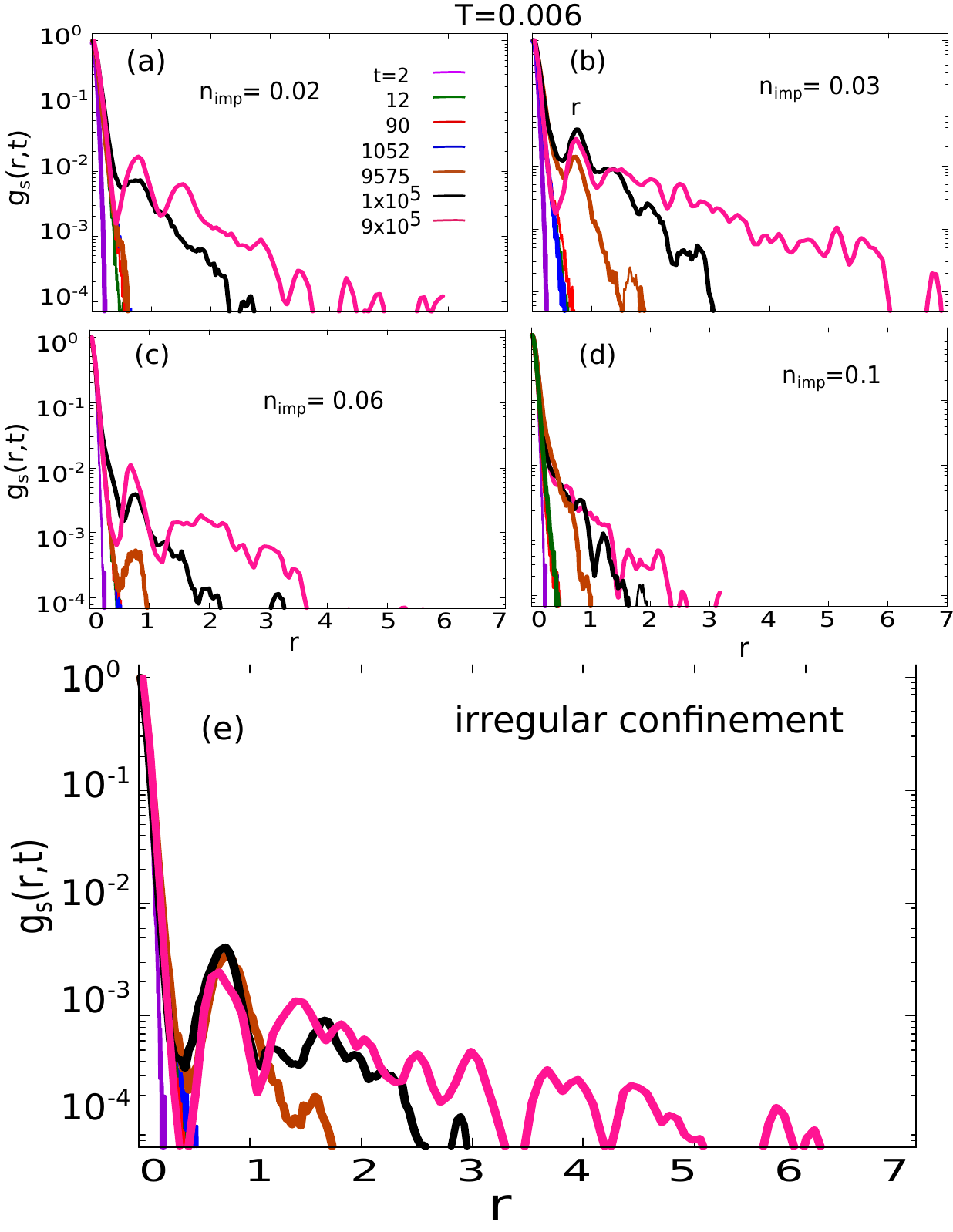}
\caption{The $r$-dependence of $G_{s}(r,t)$ are shown for different $t$. Panels (a-d) correspond to fixed $n_{\rm imp}$'s, while panel (e) represents the results from irregular trap. The behavior of $G_{s}(r,t)$ can be classified into $3$ temporal regimes (see text). For each $n_{\rm imp}$, $G_{s}(r,t)$ show non-Gaussian behavior for large $t$. Larger $n_{\rm imp}$ causes slower dynamics, so that particles diffuse up to a lesser distance as $n_{\rm imp}$ increases. The results of irregular trap are comparable only to the weak $n_{\rm imp}$, which is similar to dMSD results in Fig.~\ref{f8}. }
\label{f10}
\end{figure}

The most convenient way to explore the spatio-temporal evolution of a system is the study of the self part of the van Hove correlation function~\cite{mcdonald2006theory, PhysRev.95.249}, defined as
\begin{equation}\label{e9}
G_{s}(r,t)= \langle \frac{1}{N} \sum_{i=1}^{N} \delta(\vec{r}-(\vec{r}_{i}(t)-\vec{r}_{i}(0))) \rangle.
\end{equation}
Physically, $G_{s}(r,t)$ gives the probability that a particle has moved by a distance $r$ in time $t$, on an average. Our results for $G_{s}(r,t)$ for low temperature ($T=0.006$) are presented in Fig.~\ref{f10}, where panels (a-d) describe results for $n_{\rm imp}=0.02, 0.033, 0.06$ and $0.10$. The results for $G_{s}(r,t)$ on the irregular trap is also shown in panel (e) for comparison.

 As noted earlier\cite{PhysRevE.96.042105}, $ G_{s}(r,t) $ data can be broadly classified into three temporal regions -- a small $t$ where $ G_{s}(r) $ is gaussian, an intermediate temporal region where $ G_{s}(r)$ shows no appreciable change as a function of $t$, and a third, where $ G_{s}(r) $ features multiple peaks. Does a pinned trap show similar qualitative behaviour? Our results in Fig.~\ref{f10}(a-d) for random pinned system are qualitatively similar to above findings; however, the boundary of each of the three temporal regions depends on $n_{\rm imp}$.\\
 Having studied the Spatio-temporal correlation in our confined systems with Coulomb particles, we proceed to analyze the chaoticity and how that changes due to interaction. For this purpose, we turn to a detailed Poincare section analysis.
\section{Poincare Section Analysis}
As discussed already, the nature of motion a single particle in disordered traps could differ depending on the microscopic nature of disorder, e.g., ballistic yet chaotic motion in $V_{\rm conf}^{\rm Ir}(r)$, while diffusive~\cite{doi:10.1080/09500340210145286} in $V_{\rm conf}^{\rm Cr}(r)$ with pinning centers.
It is interesting to ask how the nature of the dynamics of a single particle in a disordered environment gets affected in the presence of other interacting particles? Can the presence of interaction among the particles make the motional signatures broadly equivalent in two differently disordered environments? 
In the subsequent sections, we address these questions by analyzing the structure of the phase space for a single particle in the two confinements. We also study the phase space structure of one particle in the presence of another Coulomb interacting particle in the above two cases. In order to understand the phase space structure for different cases, we focus on analyzing the Poincare section~\cite{strogatz2000nonlinear,hilborn2000chaos} of the trajectory of a particle obtained by integrating the Hamilton equation of motion for a given initial condition. Each time the trajectory crosses a given plane, in our case, $y = 0$ plane, with positive momentum $p_y$, the intersection coordinates $(x, p_x)$ are considered to sample the Poincare section. After a large number of intersections with the given plane, the resulting structure of the Poincare section allows us to understand the nature of the trajectory of the particle under consideration. For chaotic motion, Poincare section consists of a two-dimensional region~\cite{strogatz2000nonlinear, hilborn2000chaos, BOHIGAS199343} with randomly scattered points.\\
In the following we proceed to report our results of Poincare section analysis of the dynamics of single and two particles in our disordered confinements.

\subsection{Poincare section analysis for single-particle}

\begin{figure}[t!]
\centering
\includegraphics[width=0.46\textwidth]{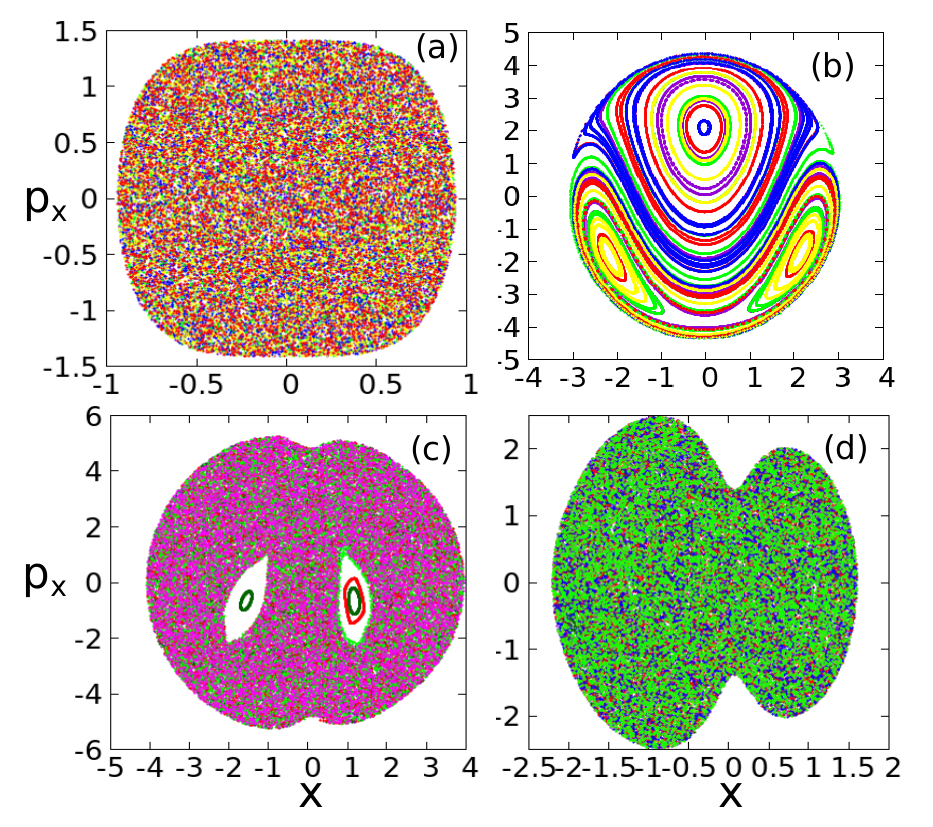}
\caption{The Poincare plane which encodes the dynamical signature of a single particle, is shown for pinned circular and irregular taps. Panel (a) depicts the chaotic nature of the single-particle dynamics in the irregular trap. Panel (b-d) shows the Poincare section of the single-particle ($E=10$) in pinned circular trap with the number of pinned particles, $n_{\rm pin}=1, 10$ and $15$ respectively. The motional signature varies from a regular (periodic) in panel (b), to a mixed motion in panel (c), to chaotic motion in panel (d). Different colors of dots are results from a different set of initial conditions for the Poincare section analysis.}
\label{f0}
\end{figure}

First, we study the motion of a single particle in the irregular trap. Since $V_{\rm conf}^{\rm Ir}(r)$ is a homogeneous function we can scale the associated Hamiltonian, and thus study of a single energy surface is sufficient to explore the dynamical features of the single particle in the system. Hence the only case of unit energy of the test particle ($E=1$) is investigated. Previous studies had shown that for $\lambda \in [0.565, 0.635]$ and $\gamma \in [0.10, 0.20]$ the dynamics of a single particle is chaotic~\cite{PhysRevLett.90.026806, BOHIGAS199343}, which is consistent with the Poincare section shown in Fig. \ref{f0}(a) where we analyze the trajectory of a single particle in $V_{\rm conf}^{\rm Ir}(r)$ for $\lambda =0.635$ and $\gamma =0.1$. As we see from Fig.~\ref{f0}(a), the Poincare section consists of a two dimensional region nearly uniformly filled with randomly placed points (different colors represent points obtained for different initial conditions); a typical signature of a chaotic motion.
 
 Next, we study the Poincare section for a single particle with pinning disorder in $V_{\rm conf}^{\rm Cr}(r)$. While $V_{\rm conf}^{\rm Cr}(r)$ is a homogeneous function, the presence of the Coulomb interaction between the test particle and the pinning centers breaks the homogeneity condition for the total Hamiltonian. As a result, the motional signature depends on the value of the total energy of the system. Thus, for the pinned system (with the number of pinning centers denoted as $n_{\rm pin}$) we consider Poincare section for different energy surfaces. Note that the dynamics of a single particle in $V_{\rm conf}^{\rm Cr}(r)$ ($n_{\rm pin}=0$) is periodic at any value of $E$. For $E=10$, we analyze the Poincare section with $n_{\rm pin}=1, 10,$ and $15$ in Fig.~\ref{f0}(b-d). For $n_{\rm pin}=1$, Poincare section consists of closed orbits implying regularity of the dynamics of the particle (see Fig.~\ref{f0}(b)). With increasing $n_{\rm pin}$ such closed orbits start to deform, and for $n_{\rm pin}=10$, we find that the most parts of the Poincare section get filled up uniformly with some elliptical orbits as shown in Fig.~\ref{f0}(c). Such smooth curves (Kolmogorov - Arnold - Moser (KAM) tori \cite{feldmeier2022introduction}) in the Poincare section correspond to regular, quasi-periodic motion while the clouds correspond to chaotic motion~\cite{strogatz2000nonlinear, hilborn2000chaos}. Thus, for $n_{\rm pin}=10$ the system exhibits mixture of chaos and order. With increasing $n_{\rm pin}$ the proportion of phase space occupied by such regular regions diminishes continually, and for $n_{\rm pin}=15$ the Poincare section gets filled up almost uniformly representing a fully chaotic dynamics of the particle for all initial conditions (see Fig.~\ref{f0}(d)). We have not noticed any invariant tori within our numerical accuracy for $n_{\rm pin}=15$. Thus, for a fixed $E (=10)$ we find that the nature of the dynamics depends on the number of pinned particles in the system.
\begin{figure}[!t]
\centering
\includegraphics[width=0.46\textwidth]{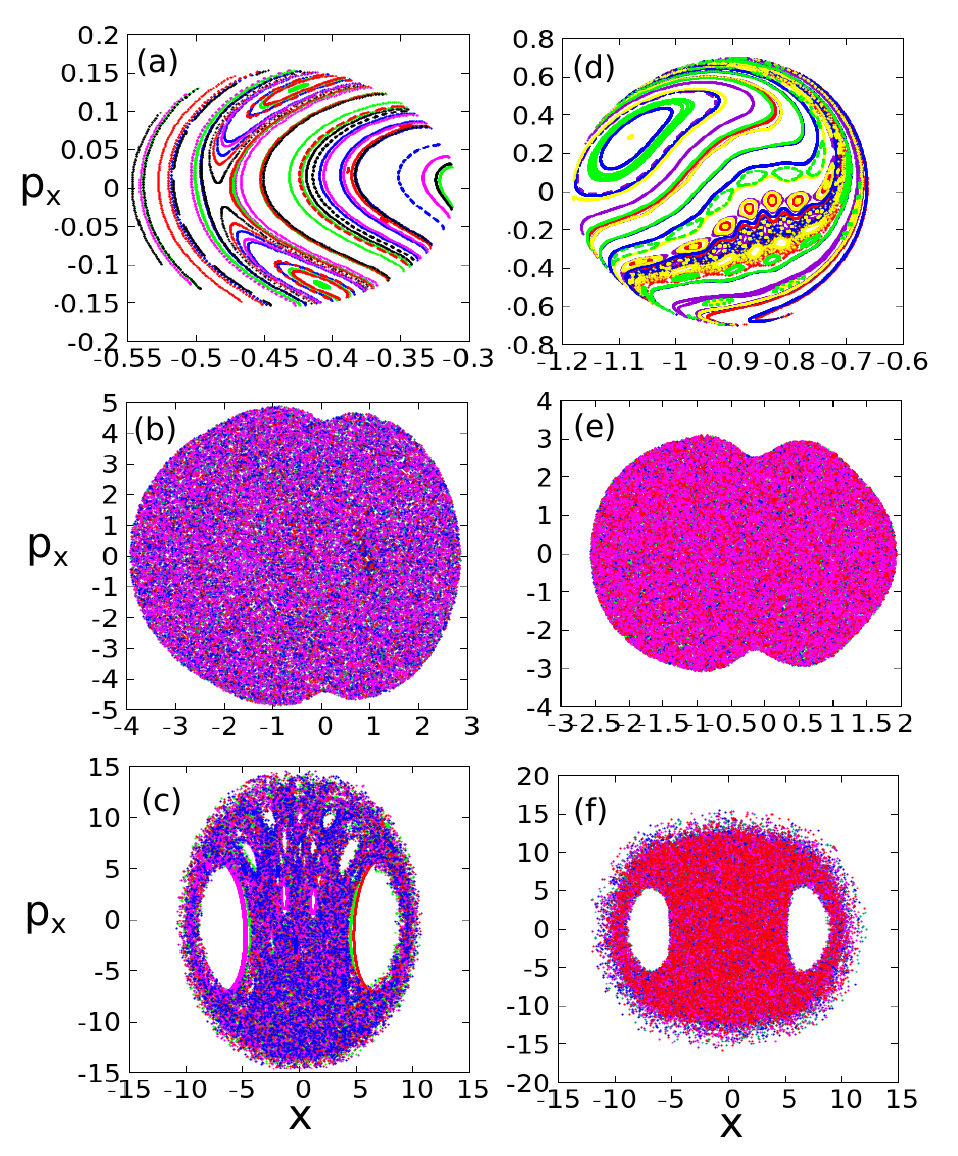}
\caption{The Poincare section of the single-particle in the circular trap with $n_{\rm pin}=15$ and $n_{\rm pin}=100$ at different energy values. Panel (a-c) shows the dynamics of the particle for $n_{\rm pin}=15$, and the corresponding energies are $E=5.7, 18.9$ and $120$, respectively. Panel (d-f) for  $n_{\rm pin}=100$ with corresponding energies $E=35, 40$ and $160$, respectively.
These results show that beyond a threshold $n_{\rm pin}$ disorder, the system exibits choaticity within an energy interval, which depends on $n_{\rm pin}$. Within this window, it is hard to distinguish the resulting diffusive single-particle motion from a chaotic dynamics.} 

\label{f12}
\end{figure}
Next, we analyze the effect of $E$ on the nature of the dynamics for fixed $n_{\rm pin}$. The Poincare section for $n_{\rm pin}=15$ with $E=5.7$ is shown in Fig.~\ref{f12}(a). It consists of elliptic and hyperbolic orbits indicating regular, quasi-periodic motion of the particle at this energy level. A generic feature of quasi-periodic dynamics is the appearance of hyperbolic and elliptic points in the Poincare section. Interestingly, when the energy is further increased by a significant amount, here $E=120$ (Fig.~\ref{f12}(c)), the Poincare section gets divided into small islands along with a large chaotic sea. Therefore, the classical motion of the particle in circular confinement with $n_{\rm pin}=15$ remains chaotic in nature for a certain range of energy. Beyond this energy range, it goes into a mixed state. We also analyze the Poincare section with $n_{\rm pin}=100$ for $E=5.7,18.9,$ and $120$. Similar to the circular system with $n_{\rm pin}=15$, $n_{\rm pin}=100$ system also passes from regular to chaotic and chaotic to mixed with an increase in energy (see  Fig.~\ref{f12}(d-f)). Thus, we find that while the nature of the dynamics of a single particle depends on the details of the disorder, there exists a broad parameter regime where a single particle exhibits qualitatively similar motional signatures in an irregular confinement and a circular confinement with pinned particles.
How does structure of the Poincare section get altered in the presence of inter-particle interaction? To address this question, we next study the Poincare section for the irregular and pinned system with two Coulomb interacting particles. 

\subsection{Poincare section analysis for two-particles system}
 
\begin{figure}[!t]
\centering
\includegraphics[width=0.46\textwidth]{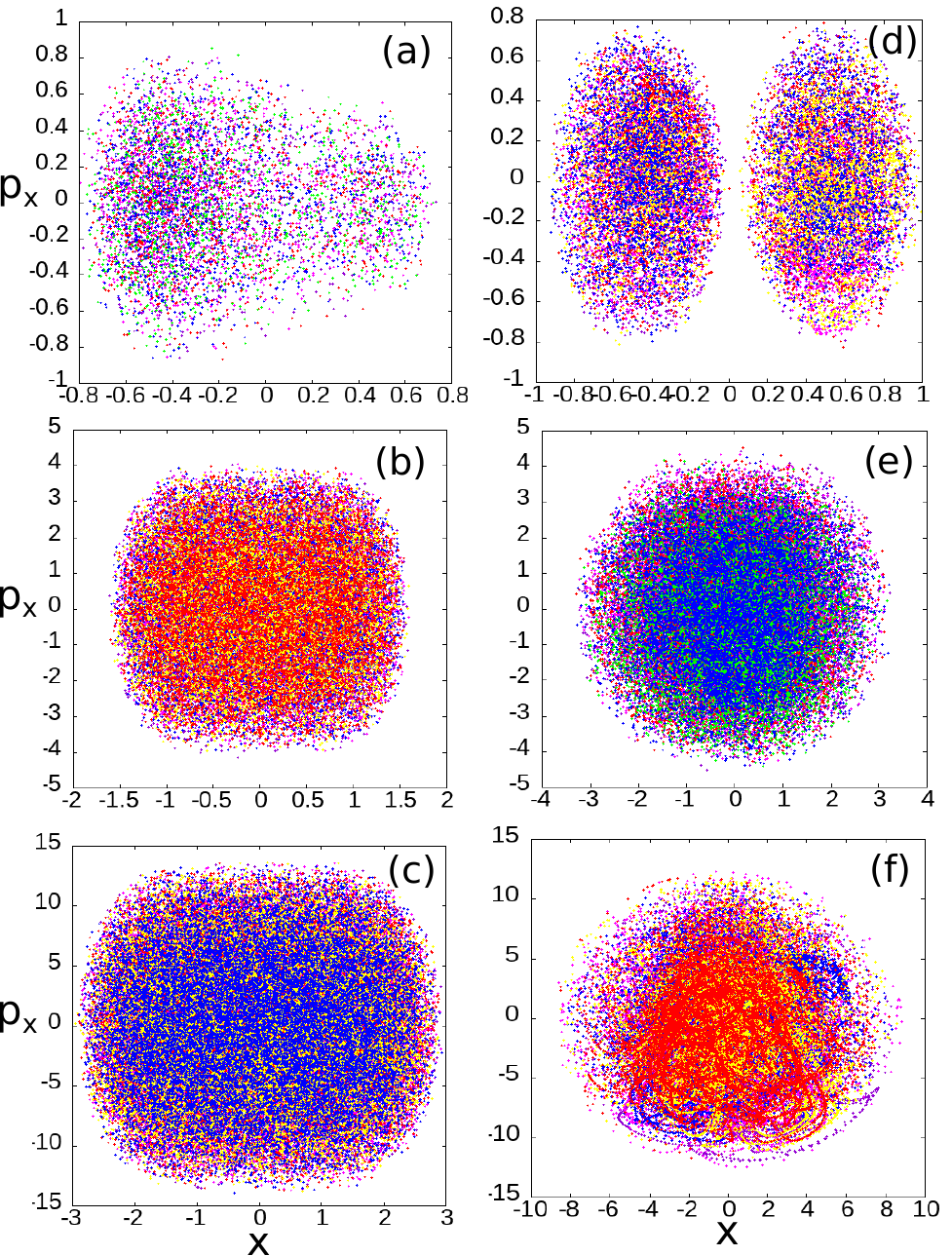}
\caption{The Poincare section for motion of $2$ particles. Panels (a-c) shows results for an irregular confinement with $E=1.5, 9$ and $100$. Panel (d-f) represents the Poincare plane for circular trap with $n_{\rm pin} =1$ and for $E=16.3, 30$ and $100$ respectively. The $2$-Coulomb particles in the $V_{\rm conf}^{\rm Ir}(r)$ continues its chaotic motion independent of $E$. In comparison, the $2$ particles in pinned system lifts the higher threshold energy to exhibit chaotic motion.}
\label{f14}
\end{figure}

 To understand the effect of interaction on the dynamics of the particles, we first analyze the Poincare section resulting from the dynamics of two Coulomb interacting particles in the irregular confinement. To obtain the Poincare section, we focus on the trajectory of any one particle and follow the same procedure as we did for a single particle. The Poincare section associated with the system of two Coulomb interacting particles confined in the irregular confinement is shown in Fig.~\ref{f14} (a-c) for $E=1.5, 9$ and $100$, respectively. We see that, like one-particle, the two-particle irregular system exhibits chaotic dynamics at any energy value. We note that the Poincare section obtained by analyzing the trajectory of the second particle exhibits qualitatively similar structure and thus, we show the Poincare section for one-particle in Fig.~\ref{f14} (a-c).
How does pinning affect the dynamics of Coulomb interacting particles? To understand that we study the Poincare section for pinned system (having $n_{\rm pin}=1)$ with two Coulomb interacting particles for different energies.

Interestingly, at low energy $(E=16.3)$ where Coulomb interaction dominates the kinetic energy, the presence of pinned particle, which interacts strongly with the two other particles, divides the Poincare section into two disjoint parts as shown in Fig.~\ref{f14}(d). 
With increasing energy particles start to overcome the Coulomb energy barrier, and at sufficiently high energy we get a uniformly filled Poincare section representing chaotic dynamics as shown in Fig.~\ref{f14} (e-f). Note that a single particle in the pinned system shows mixed dynamics for very low and high energy while it exhibits chaotic dynamics only for a given energy range (see Fig.~\ref{f12}(a-d)). In presence of Coulomb interaction we find that beyond a low-energy regime the system exhibits chaotic dynamics. Thus our analysis indicates that disorder along with inter-particle interaction can make the dynamics more chaotic, and the details of the nature of the disorder becomes less relevant.

\section{Conclusion}
Our study reveals that while qualitative physics remains more or less robust with respect to the
melting, there are quantitative differences in the details for the two different disordered systems. 
Some of our key results are - pinning disorder preserves the positional
order, while irregular confinements weaken it; the thermal crossover occurs
at nearly similar temperatures for both types of disordered confinements; signature of heterogeneous dynamics akin to glassy systems, including caging effects, emerges in both types of disorder, though the corresponding relaxation time scales are very different depending on the nature of the disorder. The Poincare section analysis reflects that the underlying dynamics of both the disordered system are chaotic. which is essentially a universal property of a disordered system.

\section{Acknowledgments}
The authors thank Chandan Dasgupta for valuable discussions. We acknowledge DIRAC computational facility of IISER Kolkata.

\bibliography{ref}

\begin{thebibliography}{69}
\expandafter\ifx\csname natexlab\endcsname\relax\def\natexlab#1{#1}\fi
\expandafter\ifx\csname bibnamefont\endcsname\relax
  \def\bibnamefont#1{#1}\fi
\expandafter\ifx\csname bibfnamefont\endcsname\relax
  \def\bibfnamefont#1{#1}\fi
\expandafter\ifx\csname citenamefont\endcsname\relax
  \def\citenamefont#1{#1}\fi
\expandafter\ifx\csname url\endcsname\relax
  \def\url#1{\texttt{#1}}\fi
\expandafter\ifx\csname urlprefix\endcsname\relax\def\urlprefix{URL }\fi
\providecommand{\bibinfo}[2]{#2}
\providecommand{\eprint}[2][]{\url{#2}}

\bibitem[{\citenamefont{Cammarota and Biroli}(2013)}]{pp1}
\bibinfo{author}{\bibfnamefont{C.}~\bibnamefont{Cammarota}} \bibnamefont{and}
  \bibinfo{author}{\bibfnamefont{G.}~\bibnamefont{Biroli}},
  \bibinfo{journal}{The Journal of Chemical Physics}
  \textbf{\bibinfo{volume}{138}}, \bibinfo{pages}{12A547}
  (\bibinfo{year}{2013}).

\bibitem[{\citenamefont{Ozawa et~al.}(2018)\citenamefont{Ozawa, Ikeda,
  Miyazaki, and Kob}}]{PhysRevLett.121.205501}
\bibinfo{author}{\bibfnamefont{M.}~\bibnamefont{Ozawa}},
  \bibinfo{author}{\bibfnamefont{A.}~\bibnamefont{Ikeda}},
  \bibinfo{author}{\bibfnamefont{K.}~\bibnamefont{Miyazaki}}, \bibnamefont{and}
  \bibinfo{author}{\bibfnamefont{W.}~\bibnamefont{Kob}},
  \bibinfo{journal}{Phys. Rev. Lett.} \textbf{\bibinfo{volume}{121}},
  \bibinfo{pages}{205501} (\bibinfo{year}{2018}),
  \urlprefix\url{https://link.aps.org/doi/10.1103/PhysRevLett.121.205501}.

\bibitem[{\citenamefont{Cammarota and Biroli}(2012)}]{ig}
\bibinfo{author}{\bibfnamefont{C.}~\bibnamefont{Cammarota}} \bibnamefont{and}
  \bibinfo{author}{\bibfnamefont{G.}~\bibnamefont{Biroli}},
  \bibinfo{journal}{Proceedings of the National Academy of Sciences}
  \textbf{\bibinfo{volume}{109}}, \bibinfo{pages}{8850} (\bibinfo{year}{2012}),
  \eprint{https://www.pnas.org/doi/pdf/10.1073/pnas.1111582109},
  \urlprefix\url{https://www.pnas.org/doi/abs/10.1073/pnas.1111582109}.

\bibitem[{\citenamefont{Brito et~al.}(2013)\citenamefont{Brito, Parisi, and
  Zamponi}}]{C3SM50998B}
\bibinfo{author}{\bibfnamefont{C.}~\bibnamefont{Brito}},
  \bibinfo{author}{\bibfnamefont{G.}~\bibnamefont{Parisi}}, \bibnamefont{and}
  \bibinfo{author}{\bibfnamefont{F.}~\bibnamefont{Zamponi}},
  \bibinfo{journal}{Soft Matter} \textbf{\bibinfo{volume}{9}},
  \bibinfo{pages}{8540} (\bibinfo{year}{2013}),
  \urlprefix\url{http://dx.doi.org/10.1039/C3SM50998B}.

\bibitem[{\citenamefont{Duan et~al.}(2021)\citenamefont{Duan, Mahault, Ma, Shi,
  and Chat\'e}}]{PhysRevLett.126.178001}
\bibinfo{author}{\bibfnamefont{Y.}~\bibnamefont{Duan}},
  \bibinfo{author}{\bibfnamefont{B.}~\bibnamefont{Mahault}},
  \bibinfo{author}{\bibfnamefont{Y.-q.} \bibnamefont{Ma}},
  \bibinfo{author}{\bibfnamefont{X.-q.} \bibnamefont{Shi}}, \bibnamefont{and}
  \bibinfo{author}{\bibfnamefont{H.}~\bibnamefont{Chat\'e}},
  \bibinfo{journal}{Phys. Rev. Lett.} \textbf{\bibinfo{volume}{126}},
  \bibinfo{pages}{178001} (\bibinfo{year}{2021}),
  \urlprefix\url{https://link.aps.org/doi/10.1103/PhysRevLett.126.178001}.

\bibitem[{\citenamefont{Angelani et~al.}(2018)\citenamefont{Angelani, Paoluzzi,
  Parisi, and Ruocco}}]{bir1}
\bibinfo{author}{\bibfnamefont{L.}~\bibnamefont{Angelani}},
  \bibinfo{author}{\bibfnamefont{M.}~\bibnamefont{Paoluzzi}},
  \bibinfo{author}{\bibfnamefont{G.}~\bibnamefont{Parisi}}, \bibnamefont{and}
  \bibinfo{author}{\bibfnamefont{G.}~\bibnamefont{Ruocco}},
  \bibinfo{journal}{Proceedings of the National Academy of Sciences}
  \textbf{\bibinfo{volume}{115}}, \bibinfo{pages}{8700} (\bibinfo{year}{2018}),
  \eprint{https://www.pnas.org/doi/pdf/10.1073/pnas.1805024115},
  \urlprefix\url{https://www.pnas.org/doi/abs/10.1073/pnas.1805024115}.

\bibitem[{\citenamefont{Karmakar and Parisi}(2013)}]{pnas}
\bibinfo{author}{\bibfnamefont{S.}~\bibnamefont{Karmakar}} \bibnamefont{and}
  \bibinfo{author}{\bibfnamefont{G.}~\bibnamefont{Parisi}},
  \bibinfo{journal}{Proceedings of the National Academy of Sciences}
  \textbf{\bibinfo{volume}{110}}, \bibinfo{pages}{2752} (\bibinfo{year}{2013}),
  \eprint{https://www.pnas.org/doi/pdf/10.1073/pnas.1222848110},
  \urlprefix\url{https://www.pnas.org/doi/abs/10.1073/pnas.1222848110}.

\bibitem[{\citenamefont{Szamel and Flenner}(2013)}]{Szamel_2013}
\bibinfo{author}{\bibfnamefont{G.}~\bibnamefont{Szamel}} \bibnamefont{and}
  \bibinfo{author}{\bibfnamefont{E.}~\bibnamefont{Flenner}},
  \bibinfo{journal}{{EPL} (Europhysics Letters)}
  \textbf{\bibinfo{volume}{101}}, \bibinfo{pages}{66005}
  (\bibinfo{year}{2013}),
  \urlprefix\url{https://doi.org/10.1209/0295-5075/101/66005}.

\bibitem[{\citenamefont{Yunker et~al.}(2010)\citenamefont{Yunker, Zhang, and
  Yodh}}]{PhysRevLett.104.015701}
\bibinfo{author}{\bibfnamefont{P.}~\bibnamefont{Yunker}},
  \bibinfo{author}{\bibfnamefont{Z.}~\bibnamefont{Zhang}}, \bibnamefont{and}
  \bibinfo{author}{\bibfnamefont{A.~G.} \bibnamefont{Yodh}},
  \bibinfo{journal}{Phys. Rev. Lett.} \textbf{\bibinfo{volume}{104}},
  \bibinfo{pages}{015701} (\bibinfo{year}{2010}),
  \urlprefix\url{https://link.aps.org/doi/10.1103/PhysRevLett.104.015701}.

\bibitem[{\citenamefont{Strandburg}(1988)}]{RevModPhys.60.161}
\bibinfo{author}{\bibfnamefont{K.~J.} \bibnamefont{Strandburg}},
  \bibinfo{journal}{Rev. Mod. Phys.} \textbf{\bibinfo{volume}{60}},
  \bibinfo{pages}{161} (\bibinfo{year}{1988}),
  \urlprefix\url{https://link.aps.org/doi/10.1103/RevModPhys.60.161}.

\bibitem[{\citenamefont{Kosterlitz and Thouless}(1972)}]{KT1}
\bibinfo{author}{\bibfnamefont{J.~M.} \bibnamefont{Kosterlitz}}
  \bibnamefont{and} \bibinfo{author}{\bibfnamefont{D.~J.}
  \bibnamefont{Thouless}}, \bibinfo{journal}{Journal of Physics C: Solid State
  Physics} \textbf{\bibinfo{volume}{5}}, \bibinfo{pages}{L124}
  (\bibinfo{year}{1972}).

\bibitem[{\citenamefont{Kosterlitz and Thouless}(1973)}]{KT2}
\bibinfo{author}{\bibfnamefont{J.~M.} \bibnamefont{Kosterlitz}}
  \bibnamefont{and} \bibinfo{author}{\bibfnamefont{D.~J.}
  \bibnamefont{Thouless}}, \bibinfo{journal}{Journal of Physics C: Solid State
  Physics} \textbf{\bibinfo{volume}{6}}, \bibinfo{pages}{1181}
  (\bibinfo{year}{1973}).

\bibitem[{\citenamefont{Halperin and Nelson}(1978)}]{HN1}
\bibinfo{author}{\bibfnamefont{B.~I.} \bibnamefont{Halperin}} \bibnamefont{and}
  \bibinfo{author}{\bibfnamefont{D.~R.} \bibnamefont{Nelson}},
  \bibinfo{journal}{Phys. Rev. Lett.} \textbf{\bibinfo{volume}{41}},
  \bibinfo{pages}{121} (\bibinfo{year}{1978}).

\bibitem[{\citenamefont{Nelson and Halperin}(1979)}]{HN2}
\bibinfo{author}{\bibfnamefont{D.~R.} \bibnamefont{Nelson}} \bibnamefont{and}
  \bibinfo{author}{\bibfnamefont{B.~I.} \bibnamefont{Halperin}},
  \bibinfo{journal}{Phys. Rev. B} \textbf{\bibinfo{volume}{19}},
  \bibinfo{pages}{2457} (\bibinfo{year}{1979}).

\bibitem[{\citenamefont{Young}(1979)}]{Young}
\bibinfo{author}{\bibfnamefont{A.~P.} \bibnamefont{Young}},
  \bibinfo{journal}{Phys. Rev. B} \textbf{\bibinfo{volume}{19}},
  \bibinfo{pages}{1855} (\bibinfo{year}{1979}).

\bibitem[{\citenamefont{Kapfer and Krauth}(2015)}]{PhysRevLett.114.035702}
\bibinfo{author}{\bibfnamefont{S.~C.} \bibnamefont{Kapfer}} \bibnamefont{and}
  \bibinfo{author}{\bibfnamefont{W.}~\bibnamefont{Krauth}},
  \bibinfo{journal}{Phys. Rev. Lett.} \textbf{\bibinfo{volume}{114}},
  \bibinfo{pages}{035702} (\bibinfo{year}{2015}),
  \urlprefix\url{https://link.aps.org/doi/10.1103/PhysRevLett.114.035702}.

\bibitem[{\citenamefont{Bernard and Krauth}(2011)}]{PhysRevLett.107.155704}
\bibinfo{author}{\bibfnamefont{E.~P.} \bibnamefont{Bernard}} \bibnamefont{and}
  \bibinfo{author}{\bibfnamefont{W.}~\bibnamefont{Krauth}},
  \bibinfo{journal}{Phys. Rev. Lett.} \textbf{\bibinfo{volume}{107}},
  \bibinfo{pages}{155704} (\bibinfo{year}{2011}),
  \urlprefix\url{https://link.aps.org/doi/10.1103/PhysRevLett.107.155704}.

\bibitem[{\citenamefont{Han et~al.}(2008)\citenamefont{Han, Ha, Alsayed, and
  Yodh}}]{PhysRevE.77.041406}
\bibinfo{author}{\bibfnamefont{Y.}~\bibnamefont{Han}},
  \bibinfo{author}{\bibfnamefont{N.~Y.} \bibnamefont{Ha}},
  \bibinfo{author}{\bibfnamefont{A.~M.} \bibnamefont{Alsayed}},
  \bibnamefont{and} \bibinfo{author}{\bibfnamefont{A.~G.} \bibnamefont{Yodh}},
  \bibinfo{journal}{Phys. Rev. E} \textbf{\bibinfo{volume}{77}},
  \bibinfo{pages}{041406} (\bibinfo{year}{2008}),
  \urlprefix\url{https://link.aps.org/doi/10.1103/PhysRevE.77.041406}.

\bibitem[{\citenamefont{Chen et~al.}(2021)\citenamefont{Chen, Tan, Wang, Zhang,
  Kosterlitz, and Ling}}]{PhysRevLett.127.018004}
\bibinfo{author}{\bibfnamefont{Y.}~\bibnamefont{Chen}},
  \bibinfo{author}{\bibfnamefont{X.}~\bibnamefont{Tan}},
  \bibinfo{author}{\bibfnamefont{H.}~\bibnamefont{Wang}},
  \bibinfo{author}{\bibfnamefont{Z.}~\bibnamefont{Zhang}},
  \bibinfo{author}{\bibfnamefont{J.~M.} \bibnamefont{Kosterlitz}},
  \bibnamefont{and} \bibinfo{author}{\bibfnamefont{X.~S.} \bibnamefont{Ling}},
  \bibinfo{journal}{Phys. Rev. Lett.} \textbf{\bibinfo{volume}{127}},
  \bibinfo{pages}{018004} (\bibinfo{year}{2021}),
  \urlprefix\url{https://link.aps.org/doi/10.1103/PhysRevLett.127.018004}.

\bibitem[{\citenamefont{Hu et~al.}(2022)\citenamefont{Hu, Zhao, and
  I}}]{PhysRevResearch.4.023116}
\bibinfo{author}{\bibfnamefont{H.-W.} \bibnamefont{Hu}},
  \bibinfo{author}{\bibfnamefont{Y.-C.} \bibnamefont{Zhao}}, \bibnamefont{and}
  \bibinfo{author}{\bibfnamefont{L.}~\bibnamefont{I}}, \bibinfo{journal}{Phys.
  Rev. Research} \textbf{\bibinfo{volume}{4}}, \bibinfo{pages}{023116}
  (\bibinfo{year}{2022}),
  \urlprefix\url{https://link.aps.org/doi/10.1103/PhysRevResearch.4.023116}.

\bibitem[{\citenamefont{Abrikosov}(1957)}]{ABRIKOSOV1957199}
\bibinfo{author}{\bibfnamefont{A.}~\bibnamefont{Abrikosov}},
  \bibinfo{journal}{Journal of Physics and Chemistry of Solids}
  \textbf{\bibinfo{volume}{2}}, \bibinfo{pages}{199} (\bibinfo{year}{1957}),
  ISSN \bibinfo{issn}{0022-3697},
  \urlprefix\url{https://www.sciencedirect.com/science/article/pii/0022369757900835}.

\bibitem[{\citenamefont{Wigner}(1934)}]{PhysRev.46.1002}
\bibinfo{author}{\bibfnamefont{E.}~\bibnamefont{Wigner}},
  \bibinfo{journal}{Phys. Rev.} \textbf{\bibinfo{volume}{46}},
  \bibinfo{pages}{1002} (\bibinfo{year}{1934}),
  \urlprefix\url{https://link.aps.org/doi/10.1103/PhysRev.46.1002}.

\bibitem[{\citenamefont{Bedanov and Peeters}(1994)}]{PhysRevB.49.2667}
\bibinfo{author}{\bibfnamefont{V.~M.} \bibnamefont{Bedanov}} \bibnamefont{and}
  \bibinfo{author}{\bibfnamefont{F.~m. c.~M.} \bibnamefont{Peeters}},
  \bibinfo{journal}{Phys. Rev. B} \textbf{\bibinfo{volume}{49}},
  \bibinfo{pages}{2667} (\bibinfo{year}{1994}),
  \urlprefix\url{https://link.aps.org/doi/10.1103/PhysRevB.49.2667}.

\bibitem[{\citenamefont{Monarkha and Syvokon}(2012)}]{rev1}
\bibinfo{author}{\bibfnamefont{Y.~P.} \bibnamefont{Monarkha}} \bibnamefont{and}
  \bibinfo{author}{\bibfnamefont{V.~E.} \bibnamefont{Syvokon}},
  \bibinfo{journal}{Low Temperature Physics} \textbf{\bibinfo{volume}{38}},
  \bibinfo{pages}{1067} (\bibinfo{year}{2012}).

\bibitem[{\citenamefont{Bonitz et~al.}(2010)\citenamefont{Bonitz, Henning, and
  Block}}]{Bonitz_2010}
\bibinfo{author}{\bibfnamefont{M.}~\bibnamefont{Bonitz}},
  \bibinfo{author}{\bibfnamefont{C.}~\bibnamefont{Henning}}, \bibnamefont{and}
  \bibinfo{author}{\bibfnamefont{D.}~\bibnamefont{Block}},
  \bibinfo{journal}{Reports on Progress in Physics}
  \textbf{\bibinfo{volume}{73}}, \bibinfo{pages}{066501}
  (\bibinfo{year}{2010}),
  \urlprefix\url{https://doi.org/10.1088/0034-4885/73/6/066501}.

\bibitem[{\citenamefont{Gann et~al.}(1979)\citenamefont{Gann, Chakravarty, and
  Chester}}]{PhysRevB.20.326}
\bibinfo{author}{\bibfnamefont{R.~C.} \bibnamefont{Gann}},
  \bibinfo{author}{\bibfnamefont{S.}~\bibnamefont{Chakravarty}},
  \bibnamefont{and} \bibinfo{author}{\bibfnamefont{G.~V.}
  \bibnamefont{Chester}}, \bibinfo{journal}{Phys. Rev. B}
  \textbf{\bibinfo{volume}{20}}, \bibinfo{pages}{326} (\bibinfo{year}{1979}),
  \urlprefix\url{https://link.aps.org/doi/10.1103/PhysRevB.20.326}.

\bibitem[{\citenamefont{B\"oning et~al.}(2008)\citenamefont{B\"oning, Filinov,
  Ludwig, Baumgartner, Bonitz, and Lozovik}}]{Bonitz_2008}
\bibinfo{author}{\bibfnamefont{J.}~\bibnamefont{B\"oning}},
  \bibinfo{author}{\bibfnamefont{A.}~\bibnamefont{Filinov}},
  \bibinfo{author}{\bibfnamefont{P.}~\bibnamefont{Ludwig}},
  \bibinfo{author}{\bibfnamefont{H.}~\bibnamefont{Baumgartner}},
  \bibinfo{author}{\bibfnamefont{M.}~\bibnamefont{Bonitz}}, \bibnamefont{and}
  \bibinfo{author}{\bibfnamefont{Y.~E.} \bibnamefont{Lozovik}},
  \bibinfo{journal}{Phys. Rev. Lett.} \textbf{\bibinfo{volume}{100}},
  \bibinfo{pages}{113401} (\bibinfo{year}{2008}).

\bibitem[{\citenamefont{Melzer et~al.}(2012)\citenamefont{Melzer, Schella,
  Miksch, Schablinkski, Block, Piel, Thomsen, Kählert, and
  Bonitz}}]{Melzer_plasma_Exp}
\bibinfo{author}{\bibfnamefont{A.}~\bibnamefont{Melzer}},
  \bibinfo{author}{\bibfnamefont{A.}~\bibnamefont{Schella}},
  \bibinfo{author}{\bibfnamefont{T.}~\bibnamefont{Miksch}},
  \bibinfo{author}{\bibfnamefont{J.}~\bibnamefont{Schablinkski}},
  \bibinfo{author}{\bibfnamefont{D.}~\bibnamefont{Block}},
  \bibinfo{author}{\bibfnamefont{A.}~\bibnamefont{Piel}},
  \bibinfo{author}{\bibfnamefont{H.}~\bibnamefont{Thomsen}},
  \bibinfo{author}{\bibfnamefont{H.}~\bibnamefont{Kählert}}, \bibnamefont{and}
  \bibinfo{author}{\bibfnamefont{M.}~\bibnamefont{Bonitz}},
  \bibinfo{journal}{Contributions to Plasma Physics}
  \textbf{\bibinfo{volume}{52}}, \bibinfo{pages}{795} (\bibinfo{year}{2012}).

\bibitem[{\citenamefont{Chu and I}(1994)}]{PhysRevLett.72.4009}
\bibinfo{author}{\bibfnamefont{J.~H.} \bibnamefont{Chu}} \bibnamefont{and}
  \bibinfo{author}{\bibfnamefont{L.}~\bibnamefont{I}}, \bibinfo{journal}{Phys.
  Rev. Lett.} \textbf{\bibinfo{volume}{72}}, \bibinfo{pages}{4009}
  (\bibinfo{year}{1994}),
  \urlprefix\url{https://link.aps.org/doi/10.1103/PhysRevLett.72.4009}.

\bibitem[{\citenamefont{Chui and Tanatar}(1995)}]{PhysRevLett.74.458}
\bibinfo{author}{\bibfnamefont{S.~T.} \bibnamefont{Chui}} \bibnamefont{and}
  \bibinfo{author}{\bibfnamefont{B.}~\bibnamefont{Tanatar}},
  \bibinfo{journal}{Phys. Rev. Lett.} \textbf{\bibinfo{volume}{74}},
  \bibinfo{pages}{458} (\bibinfo{year}{1995}),
  \urlprefix\url{https://link.aps.org/doi/10.1103/PhysRevLett.74.458}.

\bibitem[{\citenamefont{Kouwenhoven and Marcus}(1998)}]{Kouwenhoven_1998}
\bibinfo{author}{\bibfnamefont{L.}~\bibnamefont{Kouwenhoven}} \bibnamefont{and}
  \bibinfo{author}{\bibfnamefont{C.}~\bibnamefont{Marcus}},
  \bibinfo{journal}{Physics World} \textbf{\bibinfo{volume}{11}},
  \bibinfo{pages}{35} (\bibinfo{year}{1998}),
  \urlprefix\url{https://doi.org/10.1088/2058-7058/11/6/26}.

\bibitem[{\citenamefont{Bubeck et~al.}(1999)\citenamefont{Bubeck, Bechinger,
  Neser, and Leiderer}}]{PhysRevLett.82.3364}
\bibinfo{author}{\bibfnamefont{R.}~\bibnamefont{Bubeck}},
  \bibinfo{author}{\bibfnamefont{C.}~\bibnamefont{Bechinger}},
  \bibinfo{author}{\bibfnamefont{S.}~\bibnamefont{Neser}}, \bibnamefont{and}
  \bibinfo{author}{\bibfnamefont{P.}~\bibnamefont{Leiderer}},
  \bibinfo{journal}{Phys. Rev. Lett.} \textbf{\bibinfo{volume}{82}},
  \bibinfo{pages}{3364} (\bibinfo{year}{1999}),
  \urlprefix\url{https://link.aps.org/doi/10.1103/PhysRevLett.82.3364}.

\bibitem[{\citenamefont{Costa~Campos et~al.}(2013)\citenamefont{Costa~Campos,
  Apolinario, and L\"owen}}]{Simulation}
\bibinfo{author}{\bibfnamefont{L.~Q.} \bibnamefont{Costa~Campos}},
  \bibinfo{author}{\bibfnamefont{S.~W.~S.} \bibnamefont{Apolinario}},
  \bibnamefont{and} \bibinfo{author}{\bibfnamefont{H.}~\bibnamefont{L\"owen}},
  \bibinfo{journal}{Phys. Rev. E} \textbf{\bibinfo{volume}{88}},
  \bibinfo{pages}{042313} (\bibinfo{year}{2013}),
  \urlprefix\url{https://link.aps.org/doi/10.1103/PhysRevE.88.042313}.

\bibitem[{\citenamefont{Nazmitdinov et~al.}(2017)\citenamefont{Nazmitdinov,
  Puente, Cerkaski, and Pons}}]{PhysRevE.95.042603}
\bibinfo{author}{\bibfnamefont{R.~G.} \bibnamefont{Nazmitdinov}},
  \bibinfo{author}{\bibfnamefont{A.}~\bibnamefont{Puente}},
  \bibinfo{author}{\bibfnamefont{M.}~\bibnamefont{Cerkaski}}, \bibnamefont{and}
  \bibinfo{author}{\bibfnamefont{M.}~\bibnamefont{Pons}},
  \bibinfo{journal}{Phys. Rev. E} \textbf{\bibinfo{volume}{95}},
  \bibinfo{pages}{042603} (\bibinfo{year}{2017}),
  \urlprefix\url{https://link.aps.org/doi/10.1103/PhysRevE.95.042603}.

\bibitem[{\citenamefont{Yang and Lee}(1952)}]{PhysRev.87.404}
\bibinfo{author}{\bibfnamefont{C.~N.} \bibnamefont{Yang}} \bibnamefont{and}
  \bibinfo{author}{\bibfnamefont{T.~D.} \bibnamefont{Lee}},
  \bibinfo{journal}{Phys. Rev.} \textbf{\bibinfo{volume}{87}},
  \bibinfo{pages}{404} (\bibinfo{year}{1952}),
  \urlprefix\url{https://link.aps.org/doi/10.1103/PhysRev.87.404}.

\bibitem[{\citenamefont{Ghosal et~al.}(2006)\citenamefont{Ghosal,
  G{\"u}{\c{c}}l{\"u}, Umrigar, Ullmo, and Baranger}}]{Ghosal2006}
\bibinfo{author}{\bibfnamefont{A.}~\bibnamefont{Ghosal}},
  \bibinfo{author}{\bibfnamefont{A.~D.} \bibnamefont{G{\"u}{\c{c}}l{\"u}}},
  \bibinfo{author}{\bibfnamefont{C.~J.} \bibnamefont{Umrigar}},
  \bibinfo{author}{\bibfnamefont{D.}~\bibnamefont{Ullmo}}, \bibnamefont{and}
  \bibinfo{author}{\bibfnamefont{H.~U.} \bibnamefont{Baranger}},
  \bibinfo{journal}{Nature Physics} \textbf{\bibinfo{volume}{2}},
  \bibinfo{pages}{336} (\bibinfo{year}{2006}), ISSN \bibinfo{issn}{1745-2481},
  \urlprefix\url{https://doi.org/10.1038/nphys293}.

\bibitem[{\citenamefont{Beloborodov et~al.}(2003)\citenamefont{Beloborodov,
  Efetov, Lopatin, and Vinokur}}]{PhysRevLett.91.246801}
\bibinfo{author}{\bibfnamefont{I.~S.} \bibnamefont{Beloborodov}},
  \bibinfo{author}{\bibfnamefont{K.~B.} \bibnamefont{Efetov}},
  \bibinfo{author}{\bibfnamefont{A.~V.} \bibnamefont{Lopatin}},
  \bibnamefont{and} \bibinfo{author}{\bibfnamefont{V.~M.}
  \bibnamefont{Vinokur}}, \bibinfo{journal}{Phys. Rev. Lett.}
  \textbf{\bibinfo{volume}{91}}, \bibinfo{pages}{246801}
  (\bibinfo{year}{2003}),
  \urlprefix\url{https://link.aps.org/doi/10.1103/PhysRevLett.91.246801}.

\bibitem[{\citenamefont{Beloborodov et~al.}(2007)\citenamefont{Beloborodov,
  Lopatin, Vinokur, and Efetov}}]{RevModPhys.79.469}
\bibinfo{author}{\bibfnamefont{I.~S.} \bibnamefont{Beloborodov}},
  \bibinfo{author}{\bibfnamefont{A.~V.} \bibnamefont{Lopatin}},
  \bibinfo{author}{\bibfnamefont{V.~M.} \bibnamefont{Vinokur}},
  \bibnamefont{and} \bibinfo{author}{\bibfnamefont{K.~B.}
  \bibnamefont{Efetov}}, \bibinfo{journal}{Rev. Mod. Phys.}
  \textbf{\bibinfo{volume}{79}}, \bibinfo{pages}{469} (\bibinfo{year}{2007}),
  \urlprefix\url{https://link.aps.org/doi/10.1103/RevModPhys.79.469}.

\bibitem[{\citenamefont{Mirlin}(2000)}]{MIRLIN2000259}
\bibinfo{author}{\bibfnamefont{A.~D.} \bibnamefont{Mirlin}},
  \bibinfo{journal}{Physics Reports} \textbf{\bibinfo{volume}{326}},
  \bibinfo{pages}{259} (\bibinfo{year}{2000}), ISSN \bibinfo{issn}{0370-1573},
  \urlprefix\url{https://www.sciencedirect.com/science/article/pii/S0370157399000915}.

\bibitem[{\citenamefont{STÖCKMANN}(2002)}]{doi:10.1080/09500340210145286}
\bibinfo{author}{\bibfnamefont{H.-J.} \bibnamefont{STÖCKMANN}},
  \bibinfo{journal}{Journal of Modern Optics} \textbf{\bibinfo{volume}{49}},
  \bibinfo{pages}{2045} (\bibinfo{year}{2002}),
  \eprint{https://doi.org/10.1080/09500340210145286},
  \urlprefix\url{https://doi.org/10.1080/09500340210145286}.

\bibitem[{\citenamefont{Alhassid}(2000)}]{RevModPhys.72.895}
\bibinfo{author}{\bibfnamefont{Y.}~\bibnamefont{Alhassid}},
  \bibinfo{journal}{Rev. Mod. Phys.} \textbf{\bibinfo{volume}{72}},
  \bibinfo{pages}{895} (\bibinfo{year}{2000}),
  \urlprefix\url{https://link.aps.org/doi/10.1103/RevModPhys.72.895}.

\bibitem[{\citenamefont{Borgonovi et~al.}(1996)\citenamefont{Borgonovi, Casati,
  and Li}}]{PhysRevLett.77.4744}
\bibinfo{author}{\bibfnamefont{F.}~\bibnamefont{Borgonovi}},
  \bibinfo{author}{\bibfnamefont{G.}~\bibnamefont{Casati}}, \bibnamefont{and}
  \bibinfo{author}{\bibfnamefont{B.}~\bibnamefont{Li}}, \bibinfo{journal}{Phys.
  Rev. Lett.} \textbf{\bibinfo{volume}{77}}, \bibinfo{pages}{4744}
  (\bibinfo{year}{1996}),
  \urlprefix\url{https://link.aps.org/doi/10.1103/PhysRevLett.77.4744}.

\bibitem[{\citenamefont{Kim}(2003)}]{Kim_2003}
\bibinfo{author}{\bibfnamefont{K.}~\bibnamefont{Kim}},
  \bibinfo{journal}{Europhysics Letters ({EPL})} \textbf{\bibinfo{volume}{61}},
  \bibinfo{pages}{790} (\bibinfo{year}{2003}),
  \urlprefix\url{https://doi.org/10.1209/epl/i2003-00303-0}.

\bibitem[{\citenamefont{Bhattacharya and
  Ghosal}(2013)}]{bhattacharya2013melting}
\bibinfo{author}{\bibfnamefont{D.}~\bibnamefont{Bhattacharya}}
  \bibnamefont{and} \bibinfo{author}{\bibfnamefont{A.}~\bibnamefont{Ghosal}},
  \bibinfo{journal}{The European Physical Journal B}
  \textbf{\bibinfo{volume}{86}}, \bibinfo{pages}{1} (\bibinfo{year}{2013}).

\bibitem[{\citenamefont{Ash et~al.}(2017)\citenamefont{Ash, Chakrabarti, and
  Ghosal}}]{PhysRevE.96.042105}
\bibinfo{author}{\bibfnamefont{B.}~\bibnamefont{Ash}},
  \bibinfo{author}{\bibfnamefont{J.}~\bibnamefont{Chakrabarti}},
  \bibnamefont{and} \bibinfo{author}{\bibfnamefont{A.}~\bibnamefont{Ghosal}},
  \bibinfo{journal}{Phys. Rev. E} \textbf{\bibinfo{volume}{96}},
  \bibinfo{pages}{042105} (\bibinfo{year}{2017}),
  \urlprefix\url{https://link.aps.org/doi/10.1103/PhysRevE.96.042105}.

\bibitem[{\citenamefont{Ullmo et~al.}(2003)\citenamefont{Ullmo, Nagano, and
  Tomsovic}}]{PhysRevLett.90.176801}
\bibinfo{author}{\bibfnamefont{D.}~\bibnamefont{Ullmo}},
  \bibinfo{author}{\bibfnamefont{T.}~\bibnamefont{Nagano}}, \bibnamefont{and}
  \bibinfo{author}{\bibfnamefont{S.}~\bibnamefont{Tomsovic}},
  \bibinfo{journal}{Phys. Rev. Lett.} \textbf{\bibinfo{volume}{90}},
  \bibinfo{pages}{176801} (\bibinfo{year}{2003}),
  \urlprefix\url{https://link.aps.org/doi/10.1103/PhysRevLett.90.176801}.

\bibitem[{\citenamefont{Bohigas et~al.}(1993)\citenamefont{Bohigas, Tomsovic,
  and Ullmo}}]{BOHIGAS199343}
\bibinfo{author}{\bibfnamefont{O.}~\bibnamefont{Bohigas}},
  \bibinfo{author}{\bibfnamefont{S.}~\bibnamefont{Tomsovic}}, \bibnamefont{and}
  \bibinfo{author}{\bibfnamefont{D.}~\bibnamefont{Ullmo}},
  \bibinfo{journal}{Physics Reports} \textbf{\bibinfo{volume}{223}},
  \bibinfo{pages}{43} (\bibinfo{year}{1993}), ISSN \bibinfo{issn}{0370-1573},
  \urlprefix\url{https://www.sciencedirect.com/science/article/pii/037015739390109Q}.

\bibitem[{\citenamefont{Ash et~al.}(2018)\citenamefont{Ash, Dasgupta, and
  Ghosal}}]{PhysRevE.98.042134}
\bibinfo{author}{\bibfnamefont{B.}~\bibnamefont{Ash}},
  \bibinfo{author}{\bibfnamefont{C.}~\bibnamefont{Dasgupta}}, \bibnamefont{and}
  \bibinfo{author}{\bibfnamefont{A.}~\bibnamefont{Ghosal}},
  \bibinfo{journal}{Phys. Rev. E} \textbf{\bibinfo{volume}{98}},
  \bibinfo{pages}{042134} (\bibinfo{year}{2018}),
  \urlprefix\url{https://link.aps.org/doi/10.1103/PhysRevE.98.042134}.

\bibitem[{\citenamefont{Jiang et~al.}(2003)\citenamefont{Jiang, Baranger, and
  Yang}}]{PhysRevLett.90.026806}
\bibinfo{author}{\bibfnamefont{H.}~\bibnamefont{Jiang}},
  \bibinfo{author}{\bibfnamefont{H.~U.} \bibnamefont{Baranger}},
  \bibnamefont{and} \bibinfo{author}{\bibfnamefont{W.}~\bibnamefont{Yang}},
  \bibinfo{journal}{Phys. Rev. Lett.} \textbf{\bibinfo{volume}{90}},
  \bibinfo{pages}{026806} (\bibinfo{year}{2003}),
  \urlprefix\url{https://link.aps.org/doi/10.1103/PhysRevLett.90.026806}.

\bibitem[{Note1()}]{Note1}
Note1, \bibinfo{note}{a bulk system shows diverging susceptibility at a
  transition which turns into a broad peak at $T_{X}$ at the thermal crossover
  in finite systems}.

\bibitem[{\citenamefont{Frenkel and Smit}(2001)}]{frenkel2001understanding}
\bibinfo{author}{\bibfnamefont{D.}~\bibnamefont{Frenkel}} \bibnamefont{and}
  \bibinfo{author}{\bibfnamefont{B.}~\bibnamefont{Smit}},
  \emph{\bibinfo{title}{Understanding molecular simulation: from algorithms to
  applications}}, vol.~\bibinfo{volume}{1} (\bibinfo{publisher}{Elsevier},
  \bibinfo{year}{2001}).

\bibitem[{\citenamefont{Mermin and Wagner}(1966)}]{PhysRevLett.17.1133}
\bibinfo{author}{\bibfnamefont{N.~D.} \bibnamefont{Mermin}} \bibnamefont{and}
  \bibinfo{author}{\bibfnamefont{H.}~\bibnamefont{Wagner}},
  \bibinfo{journal}{Phys. Rev. Lett.} \textbf{\bibinfo{volume}{17}},
  \bibinfo{pages}{1133} (\bibinfo{year}{1966}),
  \urlprefix\url{https://link.aps.org/doi/10.1103/PhysRevLett.17.1133}.

\bibitem[{\citenamefont{Nelson}(2002)}]{nelson2002defects}
\bibinfo{author}{\bibfnamefont{D.~R.} \bibnamefont{Nelson}},
  \emph{\bibinfo{title}{Defects and geometry in condensed matter physics}}
  (\bibinfo{publisher}{Cambridge University Press}, \bibinfo{year}{2002}).

\bibitem[{\citenamefont{Tipper}(1991)}]{TIPPER1991597}
\bibinfo{author}{\bibfnamefont{J.~C.} \bibnamefont{Tipper}},
  \bibinfo{journal}{Computers \& Geosciences} \textbf{\bibinfo{volume}{17}},
  \bibinfo{pages}{597} (\bibinfo{year}{1991}), ISSN \bibinfo{issn}{0098-3004},
  \urlprefix\url{https://www.sciencedirect.com/science/article/pii/009830049190034B}.

\bibitem[{\citenamefont{Cerkaski et~al.}(2015)\citenamefont{Cerkaski,
  Nazmitdinov, and Puente}}]{PhysRevE.91.032312}
\bibinfo{author}{\bibfnamefont{M.}~\bibnamefont{Cerkaski}},
  \bibinfo{author}{\bibfnamefont{R.~G.} \bibnamefont{Nazmitdinov}},
  \bibnamefont{and} \bibinfo{author}{\bibfnamefont{A.}~\bibnamefont{Puente}},
  \bibinfo{journal}{Phys. Rev. E} \textbf{\bibinfo{volume}{91}},
  \bibinfo{pages}{032312} (\bibinfo{year}{2015}),
  \urlprefix\url{https://link.aps.org/doi/10.1103/PhysRevE.91.032312}.

\bibitem[{\citenamefont{Kong et~al.}(2003)\citenamefont{Kong, Partoens, and
  Peeters}}]{PhysRevE.67.021608}
\bibinfo{author}{\bibfnamefont{M.}~\bibnamefont{Kong}},
  \bibinfo{author}{\bibfnamefont{B.}~\bibnamefont{Partoens}}, \bibnamefont{and}
  \bibinfo{author}{\bibfnamefont{F.~M.} \bibnamefont{Peeters}},
  \bibinfo{journal}{Phys. Rev. E} \textbf{\bibinfo{volume}{67}},
  \bibinfo{pages}{021608} (\bibinfo{year}{2003}),
  \urlprefix\url{https://link.aps.org/doi/10.1103/PhysRevE.67.021608}.

\bibitem[{\citenamefont{Sun et~al.}(2016)\citenamefont{Sun, Li, Ma, and
  Zhang}}]{Sun2016}
\bibinfo{author}{\bibfnamefont{X.}~\bibnamefont{Sun}},
  \bibinfo{author}{\bibfnamefont{Y.}~\bibnamefont{Li}},
  \bibinfo{author}{\bibfnamefont{Y.}~\bibnamefont{Ma}}, \bibnamefont{and}
  \bibinfo{author}{\bibfnamefont{Z.}~\bibnamefont{Zhang}},
  \bibinfo{journal}{Scientific Reports} \textbf{\bibinfo{volume}{6}},
  \bibinfo{pages}{24056} (\bibinfo{year}{2016}), ISSN
  \bibinfo{issn}{2045-2322}, \urlprefix\url{https://doi.org/10.1038/srep24056}.

\bibitem[{\citenamefont{McDonald}(2006)}]{mcdonald2006theory}
\bibinfo{author}{\bibfnamefont{I.~R.} \bibnamefont{McDonald}},
  \emph{\bibinfo{title}{Theory of simple liquids}}
  (\bibinfo{publisher}{Academic Press}, \bibinfo{year}{2006}).

\bibitem[{\citenamefont{Pathria}(2016)}]{pathria2016statistical}
\bibinfo{author}{\bibfnamefont{R.~K.} \bibnamefont{Pathria}},
  \emph{\bibinfo{title}{Statistical mechanics}} (\bibinfo{publisher}{Elsevier},
  \bibinfo{year}{2016}).

\bibitem[{Note2()}]{Note2}
Note2, \bibinfo{note}{the smooth curve is obtained by repeated averaging of the
  falling part of $g(r)$ and subsequently, upon dividing the oscillatory
  falling part of $g(r)$ with the obtained averaged curve.}

\bibitem[{\citenamefont{Bennemann et~al.}(1999)\citenamefont{Bennemann,
  Baschnagel, Paul, and Binder}}]{BENNEMANN1999217}
\bibinfo{author}{\bibfnamefont{C.}~\bibnamefont{Bennemann}},
  \bibinfo{author}{\bibfnamefont{J.}~\bibnamefont{Baschnagel}},
  \bibinfo{author}{\bibfnamefont{W.}~\bibnamefont{Paul}}, \bibnamefont{and}
  \bibinfo{author}{\bibfnamefont{K.}~\bibnamefont{Binder}},
  \bibinfo{journal}{Computational and Theoretical Polymer Science}
  \textbf{\bibinfo{volume}{9}}, \bibinfo{pages}{217} (\bibinfo{year}{1999}),
  ISSN \bibinfo{issn}{1089-3156},
  \urlprefix\url{https://www.sciencedirect.com/science/article/pii/S1089315699000082}.

\bibitem[{\citenamefont{Donati et~al.}(1998)\citenamefont{Donati, Douglas, Kob,
  Plimpton, Poole, and Glotzer}}]{PhysRevLett.80.2338}
\bibinfo{author}{\bibfnamefont{C.}~\bibnamefont{Donati}},
  \bibinfo{author}{\bibfnamefont{J.~F.} \bibnamefont{Douglas}},
  \bibinfo{author}{\bibfnamefont{W.}~\bibnamefont{Kob}},
  \bibinfo{author}{\bibfnamefont{S.~J.} \bibnamefont{Plimpton}},
  \bibinfo{author}{\bibfnamefont{P.~H.} \bibnamefont{Poole}}, \bibnamefont{and}
  \bibinfo{author}{\bibfnamefont{S.~C.} \bibnamefont{Glotzer}},
  \bibinfo{journal}{Phys. Rev. Lett.} \textbf{\bibinfo{volume}{80}},
  \bibinfo{pages}{2338} (\bibinfo{year}{1998}),
  \urlprefix\url{https://link.aps.org/doi/10.1103/PhysRevLett.80.2338}.

\bibitem[{\citenamefont{Reis et~al.}(2007)\citenamefont{Reis, Ingale, and
  Shattuck}}]{PhysRevLett.98.188301}
\bibinfo{author}{\bibfnamefont{P.~M.} \bibnamefont{Reis}},
  \bibinfo{author}{\bibfnamefont{R.~A.} \bibnamefont{Ingale}},
  \bibnamefont{and} \bibinfo{author}{\bibfnamefont{M.~D.}
  \bibnamefont{Shattuck}}, \bibinfo{journal}{Phys. Rev. Lett.}
  \textbf{\bibinfo{volume}{98}}, \bibinfo{pages}{188301}
  (\bibinfo{year}{2007}),
  \urlprefix\url{https://link.aps.org/doi/10.1103/PhysRevLett.98.188301}.

\bibitem[{\citenamefont{Berthier}(2011{\natexlab{a}})}]{2011}
\bibinfo{author}{\bibfnamefont{L.}~\bibnamefont{Berthier}},
  \bibinfo{journal}{Physics} \textbf{\bibinfo{volume}{4}}
  (\bibinfo{year}{2011}{\natexlab{a}}), ISSN \bibinfo{issn}{1943-2879},
  \urlprefix\url{http://dx.doi.org/10.1103/Physics.4.42}.

\bibitem[{\citenamefont{Berthier}(2011{\natexlab{b}})}]{berthier2011dynamic}
\bibinfo{author}{\bibfnamefont{L.}~\bibnamefont{Berthier}},
  \bibinfo{journal}{Physics} \textbf{\bibinfo{volume}{4}}, \bibinfo{pages}{42}
  (\bibinfo{year}{2011}{\natexlab{b}}).

\bibitem[{\citenamefont{Van~Hove}(1954)}]{PhysRev.95.249}
\bibinfo{author}{\bibfnamefont{L.}~\bibnamefont{Van~Hove}},
  \bibinfo{journal}{Phys. Rev.} \textbf{\bibinfo{volume}{95}},
  \bibinfo{pages}{249} (\bibinfo{year}{1954}),
  \urlprefix\url{https://link.aps.org/doi/10.1103/PhysRev.95.249}.

\bibitem[{\citenamefont{Strogatz}(2000)}]{strogatz2000nonlinear}
\bibinfo{author}{\bibfnamefont{S.}~\bibnamefont{Strogatz}},
  \emph{\bibinfo{title}{Nonlinear dynamics and chaos westview press}}
  (\bibinfo{year}{2000}).

\bibitem[{\citenamefont{Hilborn et~al.}(2000)}]{hilborn2000chaos}
\bibinfo{author}{\bibfnamefont{R.~C.} \bibnamefont{Hilborn}}
  \bibnamefont{et~al.}, \emph{\bibinfo{title}{Chaos and nonlinear dynamics: an
  introduction for scientists and engineers}} (\bibinfo{publisher}{Oxford
  University Press on Demand}, \bibinfo{year}{2000}).

\bibitem[{\citenamefont{Feldmeier}(2022)}]{feldmeier2022introduction}
\bibinfo{author}{\bibfnamefont{A.}~\bibnamefont{Feldmeier}},
  \emph{\bibinfo{title}{Introduction to Arnold’s Proof of the
  Kolmogorov--Arnold--Moser Theorem}} (\bibinfo{publisher}{CRC Press},
  \bibinfo{year}{2022}).

\end{thebibliography}

 \bibliographystyle{apsrev}
\end{document}